\documentclass[12pt,english]{article}
\usepackage{mathptmx}

\usepackage[T1]{fontenc}
\usepackage[latin9]{inputenc}
\usepackage{geometry}
\usepackage{float}
\usepackage{amsmath}
\usepackage{amsthm}
\usepackage{amssymb}
\PassOptionsToPackage{normalem}{ulem}
\usepackage{ulem}

\makeatletter
\theoremstyle{definition}
\newtheorem*{example*}{\protect\examplename}
\theoremstyle{plain}
\newtheorem{prop}{\protect\propositionname}

\usepackage{setspace}
\usepackage[bottom]{footmisc}
\usepackage{amsfonts}
\usepackage{pdflscape}
\usepackage{amsmath}
\usepackage{datetime}

\DeclareMathOperator*{\argmax}{arg\,max}

\makeatother

\usepackage{babel}
\providecommand{\examplename}{Example}
\providecommand{\propositionname}{Proposition}

\begin{document}
\title{\noindent \textbf{Utilitarian Beliefs in Social Networks: Explaining
the Emergence of Hatred}\\
}
\author{Houda Nait El Barj, Stanford University \\
Théophile Sautory, U.C. Berkeley}
\date{{\small{}July\ 2021}}
\maketitle
\begin{abstract}
We study the dynamics of opinions in a setting where a leader has
a payoff that depends on agents' beliefs and where agents derive psychological
utility from their beliefs. Agents sample a signal that maximises
their utility and then communicate with each other through a network
formed by disjoint social groups. The leader has a choice to target
a finite set of social groups with a specific signal to influence
their beliefs and maximise his returns. Heterogeneity in agents' preferences
allows us to analyse the evolution of opinions as a dynamical system
with asymmetric forces. We apply our model to explain the emergence
of hatred and the spread of racism in a society. We show that when
information is restricted, the equilibrium level of hatred is determined
solely by the belief of the most extremist agent in the group regardless
of the inherent structure of the network. On the contrary, when information
is dense, the space is completely polarised in equilibrium with the
presence of multiple ``local truths'' which oscillate in periodic
cycles. We find that when preferences are uniformly distributed, the
equilibrium level of hatred depends solely on the value of the practical
punishment associated with holding a hate belief. Our finding suggests
that an optimal policy to reduce hatred should focus on increasing
the cost associated with holding a racist belief. \vspace{0.5in}
\pagebreak{}
\end{abstract}
\setstretch{1.5}
\begin{center}
\emph{``If societies cannot enforce the separability of the human
mind into sensitivity for what is science and for what is not science,
then we can predict that as humans evolve, the taste for truth will
disappear incrementally as it yields less pleasure'' - Friedrich
Nietzsche, Gay Science 1882.}
\par\end{center}

\section{Introduction and motivation}

Not all beliefs are born equal. Traditional models of learning often
assume that the only obstacles to the truth are agents' cognitive
capacity and the information structure. However, in some cases, the
main obstacle preventing an agent from learning the truth could be
her own self. Take the example of an employee who is being told repeatedly
by a colleague that he is doing a bad job. Our employee may choose
to ignore what his colleague is telling him because it hurts his self-image.
In fact, his colleague is sending him a truthful signal about his
work performance. After a certain period of time, the employee gets
fired for bad performance. If the truth did not hurt the employee,
he might have been more attentive to the signals sent by his colleague.
This would have potentially allowed him to improve his work performance
and avoid getting fired.

In general, knowing the truth allows agents to take optimal actions
in the real world. This is the main motivation for humans to learn
about the world in which they live in. In certain cases, knowing the
truth also yields some psychological utility, and as such this can
alter the extent to which agents learn. Knowledge of a certain topic
can create a trade-off between the psychological and the practical
utility associated with it. When learning the truth yields psychological
discomfort, we can observe that agents ``choose their own version
of reality''.

In a February 2021 article, the New York Times interviewed a number
of experts in different fields on how the Biden administration can
solve what they call our new ``reality crisis'': a world where an
increasing number of citizens chose to create ``their own version
of reality'' from hoaxes and conspiracies.\footnote{``How the Biden Administration Can Help Solve Our Reality Crisis
These steps, experts say, could prod more people to abandon the scourge
of hoaxes and lies.'' New York Times, 02/02/2021} Examples of conspiracy theories relate to whether the Covid Vaccine
will implant a tracking microchip or whether Satanic cannibalistic
pedophiles run a global child sex traffic. Conspiracy theories are
gaining in popularity and can have practical consequences for the
democracy and social stability of a country.

Indeed, various theories propogated the idea that COVID-19 was manufactured
in a Chinese lab fueling anti-Asian racism globally. Xenophobic and
violent acts targetted at the Chinese community spurged after COVID-19:
as of October 2020 more than 2,800 incidents were officially reported
in the USA.\footnote{As reported by STOP AAPI Hate.} Similarly, the
United Kingdom disclosed an increase of 21\% in anti-asian hate crimes
as compared to the previous year suggesting the pattern is not unique
to the US.\footnote{https://www.theguardian.com/world/2020/may/13/anti-asian-hate-crimes-up-21-in-uk-during-coronavirus-crisis}
In parallel, the Black Lives Matter movement raised awareness on the
ubiquitous violence and racism that impacts the global Black community.

The rise in hate crimes has opened up a social and political debate
about the roots and mechanisms of hatred. In our paper, we show that
hatred can be exactly understood as a belief that yields utility.
Starting with René Girard, the anthropological litterature argues
that in order for someone to turn hateful towards a social group,
they must be able to blame the scapegoat for some event they suffered
from. As such, degrees of hatred can be equated to strength of beliefs:
``The more I am convinced that COVID-19 has been created by the Chinese,
the more likely I am to act upon it by comitting a racist act''.
Besides, hatred can also result from network effects, where certain
individuals turn hateful the more they are surrounded with hateful
people to feel socially integrated to their circle.

In fact, hate crimes and scapegoating are not a new phenomenon. Armenians
were blamed for the decline of the Ottoman Empire, Jews were blamed
for the austerity in Germany following World War I, Tutsis were blamed
by the Hutus for Rwanda's economic crisis in the 80s...\footnote{Moise, Jean. 2014. ``The Rwandan Genocide: The True Motivations for
Mass Killings''}Not only have scapegoats emerged to take on the blame for economic
and social downturns, but various social groups have also been designated
as the cause of the appearance and spread of pandemics. Jews were
blamed for the Plague in 1347, AIDS was blamed onto the gay community
and the Chikungunya disease was blamed on the Comorean immigrants.
One could think that strong institutions, social progress, and increasing
education levels would eradicate such irrational behaviours. Yet the
Coronavirus pandemic brought scapegoating and racist acts to modern
evidence. Even in democratic and well-educated nations, the Chinese
have been blamed by some of the local population for the virus. A
natural question that arises then is how does hatred emerge and how
can we reduce it?

This paper aims to provide a general model of utilitarian beliefs
that can be applied to answer this question. We study the mechanisms
through which beliefs that yield psychological utility emerge and
spread in a social network. The recent psychological litterature has
revealed the key role of social groups and cognitive dissonance in
shaping agents' choices of beliefs. We base our model on psychological
evidence for these forces in order to better capture and explain the
dynamics of utalitarian beliefs. Our framework is based on Yariv (2002)
belief utility model where we assume that a given belief yields both
instrumental and psychological utility. Thus, when agents face a choice
of signals to sample to learn about a state of the world, they sample
whichever brings them the highest utility. In our setting, agents
have hetereogenous preferences: the relative importance of the psychological
utility in determining their choice of belief vary. This reflects
the fact that people value the truth differently. We can think of
the value of the truth as being pinned down by some psychological
cost. Some people will want to learn the truth no matter what, whereas
some other people will avoid learning the truth to keep a good self
image. We analyse a society composed of disjoint social groups where
agents want to learn a state of the world. Their belief determines
their action and the associated pay-off but also the psychological
pleasure they feel. As in DeGroot (1974), beliefs evolve as agents
communicate with other agents in their group. Through their communication,
they update their beliefs until a consensus is reached. In our framework,
the consensus reached in a social group will depend on the topology
of the network, the signals agents received as well as their underlying
preferences.

We then apply our model to explain the dynamics of hatred. We review
the psychological and anthropological litterature related to hatred.
We identify two forces: the desire to blame and the desire to belong,
that are essential in the emergence of hatred. Our general model of
utilitarian beliefs can be conveniently applied to the case of hatred,
as it encompasses these two forces. In our setting, a leader has a
political interest in blaming the cause of a downturn or pandemic
onto a social minority group. The majority of agents suffered from
this event. They hold a belief on how likely the scapegoat is responsible
for it. This belief then determines whether or not they commit a racist
act but also yields them psychological utility. If they commit a racist
act, agents receive a punishment. Individual's preferences, the structure
of social groups -their sizes, connectedness and segregation as well
as the access to information all determine the level of hatred. Our
model thus provides a rich framework through which we can study the
dynamics of hatred.

We show that when information is restricted, the level of hatred is
determined solely by the belief of the most extremist agent in the
network. However, in the presence of free and diverse source of information,
the space becomes polarised with the local consensus varying accross
groups. These consensus are driven by network effects, where some
individuals end up chosing a hate belief only after their neighbour
adopts it. On the contrary we find that when information is dense,
the equilibrium level of hatred oscillates in periodic cycles and
the space is completely polarised. This gives a greater incentive
to the leader to spread hatred. In such societies, agents can create
their own information circles to hold a belief that yields them higher
utility without inccuring the cost of social dissonance. We show that
when agents have uniform preferences, , the equilibrium level of hatred
depends uniquely on the practical punishment associated with holding
such a belief. As the number of agents in the population grows, the
population tends to satisfy these assumptions.

The primary contribution of this paper is to provide a framework guided
by psychological evidence to study the forces that shape the formation
of utilitarian beliefs. In contrast to prior work, our model studies
the dynamics of beliefs with a network fragmented into social groups
and from which agents yield utility. This allows us to study the impact
of both preferences and topological structures of communities on the
evolution of opinions. To the best of our knowledge, we are not aware
of any prior work that provides such a framework. This paper also
makes a contribution to the theoretial literature on the analysis
of dynamical systems. An advantage of our framework is that we can
define our system under asymetric forces resulting in the characterisation
of periodic equilibria. In such a setting, our model describes a population
with a polarised basis of agents, and a set of oscillating members
whose beliefs evolve. Grounded on psychological and anthropological
evidence, our model can be applied to study the dynamics of hatred
within societies. We hope that our conclusions can help inform the
design of policies to limit racism.

The remainder of the paper proceeds as follows. Section 2 presents
the model. Section 3 exposes the psychological litterature on which
our model is based, reviews the related economic litterature and lays
the groundwork for our application to hate dynamics. Consequently,
Section 4 applies the model to study the diffusion of hatred following
a major disastrous event. Section 5 concludes. The Appendix contains
the proofs for all the results.

\section{Model}

Consider a world with two possible states $\Theta=\{0,$$1\}$. Society
is composed of $N$ agents. Let $\mathcal{G}=\{G_{1},G_{2},...,G_{k},G_{k+1}\}$
be the set of all $K+1$ fixed social groups.\footnote{In our paper, we are not interested in how social groups are defined:
they could represent different political affiliations, demographics
or preferences.} We impose that agents belong to a single social group, i.e., for
all $j,k\in\{1,2,..,K+1\}$ such that$j\neq k,\hspace{1em}$$\hspace{1em}G_{j}\cap G_{k}=\emptyset$.
Agents evaluate the probability of the true state. We let $\mu_{t,i}=p_{t,i}(\theta=1)$
be the belief of agent $i$ in time $t$. At the beginning of period
$0,$ $\mu_{0,i}=0$ $\forall i$. This means that initially, all
agents believe the true state to be $\theta=0$ with certainty.

Rounds of communication take place at each period through which agents
discuss and update their beliefs. We will explain how beliefs are
updated later in this section. For now, we want to introduce the concept
of ``credibility'' and define how agents interact. Let $\phi_{i,j}$
represent the credibility of agent$j$ for agent $i$, this represents
how much agent $i$ listens and trusts agent$j$, where $\phi_{i,j}$
$\leq1$. If $\phi_{i,j}=1$, then agent $i$ believe agent $j$ to
tell the truth, and at the other extreme, $\phi_{i,j}=0$ implies
that agent $i$ does not listen to agent $j$. Define $g$ to be the
function that maps an invidiual $i$ to his social group $g$ in society,
$g:N\to\mathcal{G}$. We assume that communication happens only within
agents of the same social group, and that each agent listens to any
agent of its group, including herself, thus imposing:

\[
\phi_{i,j}=0\iff g(i)\neq g(j)
\]

There is a leader, who is not part of any social group. He has a payoff
that depends on agents' beliefs. In particular, he maximises his payoff
when all agents believe the true state to be $\theta=1$ (i.e. $\mu_{i}=1$,
$\forall i$). To influence agents' belief and increase his return,
the leader can send them a signal suggesting that the true state is
$\theta=1$ . Denote the signal sent by the leader as $s^{L}=1$.
The leader can choose which social groups to target with such signals.
All agents which are part of social groups not targetted by the leader
stick to their initial belief. This means that the leader can affect
only the beliefs of the social groups he explicitly targets.

Similarly to how agents hold grant certain credibility for each other,
agents also grant credibility to the leader. We denote by $\phi_{i}^{L}$
the credibility of the leader for agent $i$. While the credibility
of agents are fixed, we let the credibility of the leader vary. In
particular, we assume that the more social groups the leader targets
with his signals, the less credible he is. \footnote{This has a natural interpretation: when targetting a single social
group, a leader can adjust his communication to the exact preferences
of that social group in order to persuade them. As the number of social
groups he targets increases, his speech becomes less specific and
de facto less effective.} Let $\mathcal{S}\subset\mathcal{G}$ be the set of social groups
targetted by the leader. We then have that:

\[
\begin{cases}
\phi_{i,\mathcal{S}}^{L}>\phi_{i,\mathcal{S}'}^{L} & \iff|\mathcal{S}|<|\mathcal{S}'|\qquad\forall i,\mathcal{S},\mathcal{S}'\hspace{1em}g(i)\in\mathcal{S},g(i)\in\mathcal{S}'\\
\phi_{i,\mathcal{S}}^{L}=0 & \iff g(i)\notin\mathcal{S}
\end{cases}
\]

Consequently, the return of the leader will depend on the set of social
groups he targets. Let $R_{t}:\mathcal{S\to\mathbb{R}}$ be the function
that gives the return of the leader in a period $t$:

\[
R_{T}(\mathcal{S})=\sum_{i\in\mathcal{S}}\mu_{T,i}-|\mathcal{S}|c
\]

where $c$is a cost of communication. The leader maximisation problem
is thus to find a set of groups $S^{*}$ that will maximise his returns.

We now explain how agents form and update their beliefs. At the beginning
of each period $t$, agents are exposed to a single or to multiple
signals, $s_{t}\in(0,1]$ and can choose whether or not to sample
one of them. \footnote{We assume that a signal exactly equal to $0$ is impossible whereas
we allow for a signal to be exactly equal to $1$ since such a signal
is sent by the leader which we endow with communicative and manipulative
skills.}Each agent weigh a signal $s_{t}^{C}$ from source $C$ by the the
trust or accuracy they assign to the source. We denote by $\lambda_{i}^{C}$
be the trust agent $i$ gives to source $C$. Let $\hat{\mu}_{t,i}$
be agent $i$ chosen belief in period $t$, after she decided which
signal to sample (if any). Agent $i$'s chosen belief fully determines
her action in period $t$. We let $a_{t,i}$ denote the action of
agent $i$. $a_{t,i}\in\{0,1\}$ where $\hat{\mu}_{t,i}=p(a_{t,i}=1)$.

An agent chooses her belief $\hat{\mu}_{t,i}$ to maximise her per-period
utility:

\[
U(\mu_{t,i})=u(\mu_{t,i})+\sigma_{i}v(\mu_{t,i})
\]

Assume no discounting such that each util is felt in the present period
and vanishes afterwards. In this setting, the first component $u$
is the instrumental utility that the agents gets from the actions
implied by her belief. The instrumental utility function is the following
:

\[
u(\mu_{t,i})=\begin{cases}
\chi & a_{t,i}=0\\
-\chi & a_{t,i}=1
\end{cases}
\]

where $\chi>0$. The second component of our utility, $v,$ is the
psychological utility the agent gets from her belief. $\sigma$ characterises
how much an agent cares about the psychological pleasure derived from
his belief relative to its practical consequences materialised by
$\chi$. In our set up, we want the psychological utility to be maximal
when agents believe the true state of the world to be $\theta=1$
at antipodes with their instrumental utility, creating a dilemma.
An agent's belief depend on the signal they chose to sample. As such,
and to provide to provide a psychological utility in the range $[-1,1]$,
we define $v$:
\[
v(\mu_{t,i})=2\text{ \ensuremath{\lambda_{t,i}^{C}}\ensuremath{\left(s_{t,i}^{C}-\frac{1}{2}\right)}}
\]

Note that $v$is not defined when the agent chooses not to sample
a signal (i.e. when $s_{t,i}=\emptyset$ ). In such a case, we set
$v(\mu_{t,i})=0$.

Once agent have chosen their belief, rounds of communications happen
within each social group. We represent our social network as a directed
weighted graph $G(N,E(G))$ where each connected component corresponds
to a social group. Vertices correspond to agents and edges corresponds
to interactions between agents. The weight on an edge $\overrightarrow{ij}$
captures the influence of agent $i$ on agent$j$, which corresponds
to $\phi_{j,i}$ in our model and is assumed to be fixed accross periods.
\footnote{Fixing the credibility of agents can be interpreted as agents having
bounded rationality. Indeed, rational agents should update the credibility
or accuracy they give to a certain source when new information is
revealed to them.} Note that $\phi_{i,j}\neq\phi_{j,i}$ is allowed since our model
allows for asymetric influence or credibility accross agents. Let
$\Phi$ be the $n\times n$ non-negative stochastic matrix of interactions
and $\boldsymbol{\mu}_{t,k}$ be the row vector of beliefs of agents
in period $t$ after $k$ rounds of communication. The dynamics are
described by:

\[
\mu_{t,k}=\mu_{t,k-1}\Phi
\]

We then have that $\mu_{t,k+r}=\mu_{t,k}\Phi^{(r)}$ where $\Phi$
can be thought of the transition matrix of a Markov chain. We assume
that $\Phi$ is irreducible and aperiodic which allow us to apply
the martingale convergence theorem to the beliefs.\footnote{See Kemeny and Snell (1960)}
Irreducibility corresponds to the existence of a path connecting any
two vertices in the graph and implies aperiodicity when $\phi_{i,i}>0$$\forall i$
(all agents value their previous belief). Define $\omega^{*}$ as
the left row eigenvector corresponding to the unique eigenvalue equal
to $\text{1}$ for $\Phi$. After enough rounds of communication in
period $t$, beliefs converge :

\[
c_{t}=\omega^{*}\hat{\mu}_{t}
\]

Let's now illustrate the dynamics of our model with a simple example.
\begin{example*}
A society with two social groups $G_{1}$ and $G_{2}$. Only $G_{1}$
is targeted by the leader. 

There are $3$ agents in group $G_{1}$. Since $G_{2}$ is not targeted
by the leader, all agents in this social group will stick to their
prior by construction, so we are only interested in the dynamics of
the group $G_{1}$.

Let $\chi=0.4$. At the start of period $0$, they all have a belief
of exactly $0$. Now, at the begining of period $0$, they are exposed
to only one signal $s^{L}=1$. With each agent is associated a triplet
$\{\sigma,\phi^{L}\}$ . We have for agent $1$: $\{\sigma_{1}=0.1,\phi_{1}^{L}=0.2\}$,
for agent $2$: $\{\sigma_{2}=0.9,\phi_{2}^{L}=0.5\}$ and for agent
$3$: $\{\sigma_{3}=0.9,\phi_{3}^{L}=0.8\}$.

We characterise the matrix of interactions $\Phi$ in this example
to be :

\[
\Phi=\left(\begin{array}{ccc}
0.1 & 0.4 & 0.5\\
0.1 & 0.6 & 0.3\\
0.3 & 0.5 & 0.2
\end{array}\right)
\]

Let's analyse the belief decision of agent 1.

If he samples no signal, i.e. $s_{0}=\emptyset$, then he stays at
his prior which is $\mu_{o}=0$ and so his utility is:

\begin{align*}
U(\hat{\mu}_{0}|s_{o}=\emptyset)= & -\chi\cdot\mu_{0}+\chi\cdot(1-\mu_{0})+\sigma_{1}\cdot0\\
= & 0\cdot(-0.4)+0.4\cdot1+0.1\cdot0\\
= & 0.4
\end{align*}

If he samples $s_{0}=1$, then his chosen belief is $\hat{\mu}_{o}=\phi_{1}^{L}\cdot s^{L}+(1-\phi_{1}^{L})\cdot\mu_{o}=0.2\cdot1+0.8\cdot0=0.2$
and so his utility is

\begin{align*}
U(\hat{\mu}_{0}|s_{o}=1)= & -\chi\cdot\mu_{0}+\chi\cdot(1-\mu_{0})+\sigma_{1}\cdot2\text{ \ensuremath{\phi_{1}^{L}}\ensuremath{\left(s^{L}-\frac{1}{2}\right)}}\\
= & 0.2\cdot(-0.4)+0.4\cdot0.8+0.1\cdot0.1\cdot1\\
= & 0.25
\end{align*}

In this case, for agent 1, $U(\hat{\mu}_{0}|s_{o}=\emptyset)>U(\hat{\mu}_{0}|s_{o}=1)$,
and as such, he will chose not to sample the signal and stick to his
prior. Thus, his chosen belief at the begining of period $0$, is
$\hat{\mu}_{0,1}=0$. Applying the same logic to agent $2$ and $3$,
we get $\hat{\mu}_{0,2}=0.5$ and $\hat{\mu}_{0,3}=0.8$. Thus, our
row vector of initial chosen belief in period $0$ is :

\[
\hat{\mu}_{0}=(\begin{array}{ccc}
0 & 0.5 & 0.8\end{array})
\]

After one round of communication, the row vector of beliefs is given
by :

\begin{align*}
\mu_{0,1}= & \hat{\mu}_{0}\cdot\Phi\\
= & (\begin{array}{ccc}
0 & 0.5 & 0.8\end{array})\cdot\left(\begin{array}{ccc}
0.1 & 0.4 & 0.5\\
0.1 & 0.6 & 0.3\\
0.3 & 0.5 & 0.2
\end{array}\right)\\
= & (\begin{array}{ccc}
0.29 & 0.70 & 0.30\end{array})
\end{align*}

After $6$ rounds of communication, beliefs eventually converge, $\mu_{0,6}=(\begin{array}{ccc}
0.51 & 0.51 & 0.51\end{array})$. Thus, $c_{0}=0.51$ and the period $0$ ends. Period $1$ begins,
and all agents of $G_{1}$ initially hold a belief $\mu_{1}=0.51$.
Then, the same steps described for period $0$ happen and similarly
for all future periods.
\end{example*}

\section{Psychological evidence and related litterature}

In our model, we want to capture the idea that the dynamics of learning
are singular in presence of utilitarian beliefs. In particular they
depend on how much weight agents place on the psychological component
of their belief. Furthermore, as humans are social beings who communicate
with one another, the dynamics of opinions should be studied from
a social network framework. This becomes even more crucial in the
presence of utilitarian beliefs. Indeed, we will show in Section $4$
that network effects can create incentives for individuals to adopt
certain beliefs that they would not have adopted with different neighbours.
Consequently, in this section, we want to provide psychological evidence
for the forces that shape the dynamics of utilitarian beliefs. We
review the psychological litterature associated with scapegoating
to show how hatred can be understood as a belief that yields utility.
This will allow us to apply our model to the dynamics of racism in
Section $4$.

In his seminal book \textquotedblleft Le bouc émisssaire\textquotedblright ,
French anthropologist René Girard establishes a theory where scapegoating
is defined as a collective phenomenon that emerges when individuals
have been experiencing an endemic and unconscious violence. By being
able to gather \textquotedblleft all against one\textquotedblright ,
the scapegoat allows for violence to be released and expressed in
society. Under his theory, the choice of the scapegoat is arbitrary,
as it only serves the purpose of an external outlet through which
the majority can get deliverance from their internal anxiety. Consequently,
moments of scapegoating in History should be recurring, as they represent
a collective crisis that rises in intensity until the internalized
violence culminates and must be released. While the designation of
the scapegoat is independent of any cause, the targetted group must
fit certain criteria. First, the scapegoat must be a minority group
that can clearly distinguished from the majority. Second, the majority
must believe that the scapegoat is somewhat responsible for the misfortune
they are accusing them of. Finally, the scapegoat must present some
extreme quality such as wealth, beauty, vice, or weakness.

René Girard\textquoteright s theory is closely related to Freud\textquoteright s
notion of projection which he defines as the \textquotedblleft self
reproach repressed by erecting defensive symptom of distrust of other
people. In this way, the subject withdraws his acknowledgement of
the self-reproach\textquotedblright (Freud, Further Remarks on the
Neuro-Psychoses Of Defence 1896b) . In Freudian psychoanalytic theory,
projection is an unconscious defense mechanism through which agents
get relief by displacing their reproach onto someone else and can
do so \textquotedblleft without any consideration for reality\textquotedblright .
Doing so allows the agent to reduce cognitive dissonance that could
result from discovering unflattering aspects of one\textquoteright s
image or reality such as the notion of guilt (Freud, Fragment of An
Analysis Of A Case Of Hysteria 1905).

Building up on Freud\textquoteright s ideas, Dollard et al. (1939)
theorize the emergence of scapegoats as the result of society's frustration
following a specific event. Individuals cannot target their violence
at the original source (either because it is absent or non accessible).
Consequently, the aggression is displaced towards an easy target,
which usually happens to be a minority group. Providing evidence for
the frustration-aggression hypothesis of Dollard, Hovland and Sears
(1940) show that incidence of lynchings of Black people in Southern
states is positively correlated with bad economic indicators. Amongst
other works, Douglas (1995) characterizes scapegoating as a strategy
used by agents to minimize feelings of guilt over the outcome of a
negative event for which they could be responsible. Related to this
theory, Rothschild et al. (2012) posit that a scapegoat allows for
the majority to keep their moral integrity by displacing guilt. In
three studies, they show that a negative outcome that could be linked
to one\textquoteright s own actions or whose source is unknown increase
scapegoating. In their theory, two fundamental forces lead to scapegoating
: the need to perceive oneself as morally valuable and the need to
perceive oneself as having control over one\textquoteright s environment.\footnote{Glick (2002) and Glick (2005) also develop a theory where scapegoating
is attractive as it allows individuals to make sense of negative outcomes
without a clear cause in a simple way, which restores the desire for
control over their environment, as they can now simply punish the
scapegoat.}

Common to all these theories is the desire to blame a social group
to get a sense of relief from the potential responsibility agents
could have in their own misfortune. The need for blame as a fundamental
process in social cognition has been documented in the psychology
litterature. Beardsley (1970) characterizes blame as \textquotedblleft a
power and poignancy for human life unparalleled by other moral concepts\textquotedblright .
While blame allows agent to project guilt onto a chosen scapegoat,
they must be able to justify that the victim deserves her mistreatment.
(McKenna, 2012). Cikara, Botvinick and Fiske (2019) show that the
more an individual perceive themselves as threatened by the minority,
the more pleasure they get from punishing them.

Malle, Guglielmo and Monroe (2014), propose a theory of blame in which
an ordering of criteria (event detection, agent causality, intentionality,
obligation, reasons and capacity) defines degrees of blame. The human
desire for meaning, and avoidance of uncertainty as well as avoidance
of the responsibility for a negative outcome all justify the desire
for blame (Malle and Knobe 1997, Wong and Weiner 1981, Hilton 2007).
Experimental evidence for blame has been provided in Gurdal, Miller
and Rustichini (2013) where principals blame agents for the negative
outcome of a lottery even though they are aware it is a random event.

Another essential force of scapegoating is operated via the collective
power of social groups versus the targeted minority. An essential
criterion is the ability of the blamers to self-dissociate from the
blamed, which must hence stem from categorization of social groups
into in-groups (the blamers) and the outgroups (the scapegoat).\footnote{ See L. Z. Tiedens \& C. W. Leach (Eds.). Studies in emotion and social
interaction. The social life of emotions (p. 314\textendash 335)}.Tajfel et al. (1971) show that more defined social categorization
allows for increased discriminative behaviors with more \textquotedblleft ingroup
favoritism\textquotedblright{} and more \textquotedblleft outgroup
prejudice\textquotedblright . Tajfel (1981) and Chen and Li (2009)
show that agents usually seek to maximize the difference between salient
in-groups and outgroups. Categorization of society into social groups
operates as the necessary structure through which differential treatments
of oneself and the other can happen, so long as they can be differentiated.
Therefore, the more distinct and distanced the scapegoat in society
relative to other groups, the easier it is for agents to direct hate
and blame at them.

At the heart of social groups lies the human need for belonging. From
an evolutionary perspective, belonging could represent a survival
advantage since groups can better hunt, protect themselves from predators,
and find potential mates to reproduce. (Mangel and Clark 1985). Baumesteir
and Leary (1995) define the need to belong as an \textquotedblleft evolutionary
selection that guide individual human beings into social groups and
lasting relationships. These mechanisms would presumably include a
tendency to orient towards other members of the species, a tendency
to experience affective distress when deprived of social contact or
relationships, and a tendency to feel pleasure or positive affect
from social contact and relatedness\textquotedblright . Baumesteir
and Wotman (1992), show that belonging to a social group is usually
associated with positive emotions such as happiness and safety, while
Baumesteir and Tice (1990), Leary and Tambor (1993), Argyle (1987)
and Myers (1992) all show that social exclusion, and a sense of isolation
as well as being deprived of relationships is associated with higher
depression, anxiety and grief. Kiecolt-Glaser et al. (1984) found
that loneliness was associated with decrease in immunocompetence and
increase in urinary cortisol respectively. Lynch (1977) summarises
evidence from many studies to conclude that \textquotedblleft U.S.
mortality rates for all causes of death are consistently higher for
divorced, single, and widowed individuals\textquotedbl{} than for
married individuals\textquotedblright .

Social groups define their own social norms, which are a set of implicit
rules that categorize what is acceptable within the group. Deviation
from one\textquoteright s group social norms has been shown to lead
to communication to produce conformity and eventually loss of social
status (Festinger 1950, and Schachter 1951). Since belonging is a
fundamental need for humans, the above imply that agents would not
in general not want to deviate from their group\textquoteright s accepted
beliefs at a given time.

The structure of social groups has a dual impact on the formation
of hatred: at the scapegoat level, it allows individual to dissociate
from that particular group and to direct the hatred at them. Further,
when social groups are rigid and static, agents feel more pressured
to adhere to the average group belief as they would not be integrated
in society otherwise. If a belief forms in a given group and the agent
would not initially choose to believe in it, she might feel pressured
to do so in order to be integrated in her own social group. However,
this pressure lessens when the agent has the possibility to belong
to a different social group if she expresses different views than
his initial group majority. Therefore, in our model, agents' desire
to belonging is an essential force in the formation of the hate belief
that is shaped by the topology of social groups.

The other essential force that we will capture is the desire for self-esteem.
This implies that agents are more inclined into believing a specific
group is responsible for the bad economic environment when they have
themselves suffered from it. Agents blame the scapegoat following
an unconscious desire to vengeance, holding them responsible for a
global event that causes them distress.\footnote{Galofré-Vilà et al. (2017) find that between 1930 and 1933, German
districts most impacted by radical austerity measured are associated
with higher vote shares for the Nazi party. https://www.nber.org/system/files/working\_papers/w24106/revisions/w24106.rev0.pdf} Even if they do not feel attacked or threatened by the scapegoat,
holding them responsible allows hatred to operate as a mean of projection
of distress into a clear recipient object. This mechanism provides
agents with relief and a sense of certainty.

To our knowledge, the economic literature on scapegoating is limited
to the work of Glaeser (2005). He uses a model of supply and demand
to derive an equilibrium level of hatred in an economy where factors
such as desire for vengeance, time spent listening to atrocities stories
about the minority as well as costs from reduced economic interactions
impact the demand side while the desire to complement policies from
a political candidate, financial resources of actors and homogenous
party platforms determine the supply side. Our work is closely related
to Glaeser (2005) since we both characterise hatred as being initially
sparked by a leader who has some political gain in blaming a minority
and where agents have some utility in feeling hateful. However, our
model uses a new framework of non-Bayesian learning where beliefs
enter the utility of agents. This allows us to study how both the
interactions within social groups and selfish preferences impact the
level of hatred. Furthermore, in our model, we are able to characterise
how some individuals are incited to hatred only because of their neighbours
being hateful. In Glaeser (2005) agents become haters as a function
of the amount of time spent hearing horrible stories, whereas in our
model, hatred is a belief that varies with the information access
and the preferences of each agent. In our setting, the macroscopical
level of hatred is determined globally by the structure of social
groups. On the contrary to Glaeser (2005), limiting factors take account
of the propensity through which individuals update their belief with
the available information and through communication.

Our work is also closely related to the economic literature on psychological
expected utility whereby beliefs about a given state impact the utility
directly. Laying the grounds for our model are Yariv, (2001); Koszegi,
(2003), Oster, Shoulson and Dorsey (2014) and Caplin and Leahy (2001).
In those models, agents actions depend on their beliefs and are the
result of a utility-maximisation problem. These beliefs yield ego-utility
which interferes with the optimal actions being taken. On the other
hand, our paper also relates to the literature on non-bayesian updating.
The closest works to ours in the questions asked are Golub and Jackson
(2010), Acemoglu et al. (2008) and De Marzo, Vayanos, and Zwiebel
(2003). Other relevant paper to our work are Banerjee (1992); Acemoglu
and Ozdaglar (2010), Bikhchandani, Hirshleifer, and Welch (1992);
Ellison and Fudenberg (1993), Acemoglu Chernozhukov, and Yildiz (2016),
Banerjee and Fudenberg (2004) Mossel, Sly and Tamuz (2015), Molavi,
Tahbaz-Salehi and Jadbabaie (2018) and Reshidi (2020).

\section{Application of our model to the dynamics of hatred}

We can now aply our model to understand how hatred emerges and spread
within a society.

Consider a society initially prosperous, until a major harsh economic
or social event happens that touches a significant share of the population.\footnote{Examples of such events can be found in history. For instance, following
World War I, Germany was held responsible for the damages caused and
had to repay an amount of 67.8 billion goldmarks, which led to a period
of austerity that touched in particular middle and lower classes.
Another modern example could be the pandemic of the Coronavirus that
led to more than 300,000 deaths in the USA.}This cause-event is disastrous at a large scale: not only does it
impact the life of many agents but the impact itself is consequential
for each agent.

Following such a large-scale event, no clear responsible can be unequivocally
identified. This represents a threat for the political leader\footnote{In our model we assume the existence of a unique leader interested
in spreading hatred. Extensions with multiple competing political
parties are left for future research.} to be blamed. Targetting the blame at a specific group represents
a political advantage where he can win some electorate. As argued
in section $3$, the chosen scapegoat group must meet two criteria
: they must be a minority to provide a political advantage to the
leader (otherwise he would loose more electorate than he can potentially
win) and the majority should be able to make a link between the current
event and the scapegoat group.\footnote{For instance, in the 1920s in Germany, the majority of middle and
lower classes lived under dire economic conditions due to the austerity
imposed while Jews were usually in the upper class and hence very
wealthy. Thus Hitler could introduce the idea that since Jews were
untouched by austerity they must have wanted it, and are hence responsible
for Germany's misery. In a similar way, Trump introduces the idea
the Chinese people are responsible for the Coronavirus pandemic since
the virus originated in China.}

Consider a world with two possible states , $\Theta=\{0,$$1\}$ with
$\theta=1$ corresponding to the event ``The scapegoat is responsible
for the event and should be punished'' and $\theta=0$ corresponding
to the event ``The scapegoat is not responsible''. Let $\mu_{t,i}=p_{t,i}(\theta=1)$
be the belief of agent $i$ in time $t$. Initially, at the onstart
of period $0$, $\mu_{0,i}=0$ $\forall i$, essentially representing
the fact that racism is not ``innate'' since we are interested in
studying its emergence and dynamics.

There are $K+1$ social groups in our society, assume that the group
$G_{K+1}$ is the scapegoat. It is a minority group associated with
the global event. Then, the leader has a choice of which subset $\mathcal{S\subseteq}\mathcal{\mathcal{G}\setminus}G_{k+1}$
to target with hateful signals about the scapegoat to incite hatred.

\subsection{A dictator world}

First, let's study the dynamics of hatred in a dictator world, where
information is restricted to the one provided by the leader.

Consider a group in $\mathcal{\mathcal{G}\setminus}G_{k+1}$ and assume
that they are targetted by the leader. We know that initially all
agents start with $\mu_{0,i}=0$, and at each period they receive
a signal $s_{t,i}=s^{L}=1,\forall i,t$ and $\lambda_{t,i}=\phi_{i}^{L},\forall t$.
Thus the choice of agents when receiving this signal is either to
sample it, or ignore it and stick to their prior. Since initially
$\mu_{i,-1}=0$ $\forall i$, agents must gain a lot of blaming-utility
to choose to adopt the leader's belief. There are huge initial costs
of holding a hate belief. \footnote{An interesting empirical question would be whether $\phi^{L}$ and
$\sigma$ are independent. Is the credibility or accuracy an agent
assigns to another dependent on how much they want to believe in their
words. Or said differently, do we tend to take as accurate statements
that we desire to believe in?} Thus after being exposed to the leader in period $0$, the beliefs
of agents are :

\[
\begin{cases}
\hat{\mu}_{0,i}=\phi_{i,L} & \text{if }\sigma_{i}>2X\\
\hat{\mu}_{0,i}=0 & \text{if }\sigma_{i}\leq2X
\end{cases}
\]

Then, after choosing their belief, they communicate, and they do so
at each new period. We then have the following result:
\begin{prop}
Equilibrium level of hatred

At finite horizons, the network topology matters, in particular, within
one round of communication we can guarantee a minimum consensus belief
at under the following condition on the network topology. We have
$\omega_{j}^{*}\leq\omega_{k}^{*}$ for every $\sigma_{j}\geq\sigma_{k}$
and $\phi_{j}^{L}<\phi_{k}^{L}$ . On the contrary, at infinite horizon,
the network topology is irrelevant. Provided there is a single agent
with $\sigma_{i}>2X$, then $lim_{k\to\infty}c_{k}=\max\{\phi_{i}^{L}\}_{i:\sigma_{i}>\frac{2X(\phi_{i}^{L}-c_{q})}{\phi_{i}^{L}},\text{for some q}}$.
In particular, if there a single agent $k$ with $\sigma_{k}>2X$
and $\phi_{k}^{L}=\max\{\phi_{i}^{L}\}_{i\in N}$ then $lim_{k\to\infty}c_{k}=\phi_{k}^{L}$
regardless of the network topology.
\end{prop}
This proposition reveals an important dynamic of a utilitarian belief.
At the finite horizons, for a given sequence of initially chosen beliefs,
as determined by $\sigma$, agents with lowest beliefs on hate should
be those who have the highest social influence on the network. Under
the dynamics of our system, the finite horizon consensus belief is
a convex combination of the leader's credibility for agents. Some
agents are forced into hatred due to the communication round and convergence
of opinions without it ever being the result of the influence of the
leader on them. Those agents can be interpreted as individuals who
do not have much incentive to develop hatred and the only way their
individual belief is contributing into the final hate consensus level
is via them adopting their group belief. On the reverse, agents with
high incentive to hatred make an individual choice to become hateful
and their adopting of the leader's belief has a direct impact on the
final belief. But at infinite horizon, the steady-state level of hatred
is independent of the network topology and at equilibirum, the leader
targets all agents in the majority group, and they all have belief
$\mu_{\infty}=\max\{\phi_{i}^{L}\}_{i:\sigma_{i}>\frac{2X(\phi_{i}^{L}-c_{q})}{\phi_{i}^{L}},\text{for some q}}$
and commit a racist act with probability $\mu_{\infty}$. The intuition
for infinite horizon equilibrium is the follow. At the beginning of
each period, agents can adopt the leader's belief weighted by his
credibility whenever it yields enough utility. Utility is increasing
in the probability that the true state is $\theta=1$, where more
agents adopt the leader's belief giving other agents an incentive
to adopt it as it establishes social norms in their society. Said
differently, the people with the highest utility to blame will be
the first to adopt the leader's belief. Even though preferences of
agents and the punishment for holding a hate belief never change,
as the consensus increases more agent have an incentive to adopt the
leader's belief as it becomes the norm. This is exactly because of
the group effect that we described in section $3$.

Given these dynamics, we are now interested in the decision of the
leader. Who will he target with hateful messages to spark hatred?

Consequently, the leader has a choice of choosing which groups in
$S=\{G_{1},G_{2},...,G_{k}\}$ to target with hateful messages in
order to maximise his returns as defined in Section $2$.

Assume that the leader cares about the level of hatred in a future
period $r$ (for example $r$ could be the election period) but sends
hateful messages in all periods $0$ to $r$.

We can thus redefine our leader maximisation problem using our assumption
of distinct groups :

\begin{align*}
\max_{i}\sum_{i\in S}\mu_{i}-|S|c\hspace{1em} & \iff\max_{G\in S}c_{k,G}|G|-|S|c\hspace{1em}\hspace{1em}\\
 & \text{{subject\hspace{1em}to\hspace{1em}}\ensuremath{\begin{cases}
 \phi_{i,\mathcal{S}}^{L}>\phi_{i,\mathcal{S}'}^{L}\hspace{1em}\iff|\mathcal{S}|<|\mathcal{S}'|  &  \hspace{1em}\hspace{1em}\forall i,\mathcal{S},\mathcal{S}'\hspace{1em}g(i)\in\mathcal{S},g(i)\in\mathcal{S}'\\
 \phi_{i,\mathcal{S}}^{L}=0\hspace{1em}\hspace{1em}\iff g(i)\notin\mathcal{S} 
\end{cases}}}\\
\end{align*}

where $g$ is the mapping function defined in section $2.$

In particular, we let $\phi_{i,S}^{L}=\max\{0,\phi_{i}^{L}-\frac{N_{s}}{K}\}$
where $N_{s}=|\cup_{G\in S}G|$ and $K=|G_{1}\cup...\cup G_{k}|$.
Then his return $R:2^{S}\to\mathbb{R}$, is given by $R(S^{'})=c_{k,G}|G|-|S^{'}|c$
for $G\in S^{'}$ and $S^{'}\subseteq S$. In the below, we will assume
that $c=0$, representing the fact that nowadays targetting individuals
with messages can be done at virtually no cost via social platforms
such as Twitter or Facebook. This assumption does not alter the qualitative
or comparative nature of our results. Adding a cost reduces the incentive
for the leader homogenously accross all groups. We will be looking
at a period $r$where all groups are in equilibrium consensus as defined
in the above section.
\begin{prop}
In a dictator world, where the leader can have full control of the
signals sent to individuals, if $\max_{i\in G}\phi_{i}^{L}$ is equal
in all groups, then the choice of the leader is simply determined
by the size of the groups. Let $S^{*}$ denote the set of social groups
maximising the leader's returns, which may not be unique, then $S^{*}$
is chosen such that $|S^{*}|=\phi^{L,\max}\cdot\frac{K}{2}$. Under
heterogenous preferences, and equal group sizes, $|g|=N\hspace{1em}\forall g,$
then the leader targets all group $g$ with $\phi_{g}>\frac{N}{K}$.
Finally, if social groups are not heterogenous in their preferences
and sizes, then the leader return maximising set must balance off
adverse effects of group sizes. In fact if $S^{*}$ is a return-maximising
set of groups, then: all omitted groups $G_{m}$ $\notin S^{*}$ are
such that $\phi_{m}<\frac{|G_{m}|}{K}+2\sum_{s\in S^{*}}\frac{|g_{s}|}{K}$
and all included groups $G_{s}$ $\in S^{*}$ verify $\phi_{s}>\frac{|G_{s}|}{K}+2\sum_{l\in S^{*},l\neq s}\frac{|g_{l}|}{K}$.
\end{prop}
Proposition $2$ calls attention to an important fact: the size of
a group is at the same time a curse and an advantange for the leader.
The bigger the group, the greater the number of people that turn hateful
if this group is targetted by the leader, which increases his returns.
However, the bigger the group, the more it represents an absolute
advantage to the leader so the more he has to adapt his communication
to fit the preferences of this group. Consequently he is less credible
for all groups, thus reducing back his returns. As such, when groups
all have a similar extreme individual (i.e. as an individual whose
credibility for the leader is maximal and similar in all groups) then
the choice of the leader simply comes back to selecting social groups
such that his choice is half the population size hence balancing the
trade off. However, when groups are very heterogenous in the agent
who has maximum credibility, then the leader has to select which groups
to target individually where both $(\phi_{G}^{L,\max},|G|)$ matter.
Intuitively, he wants to target groups with the highest possible $\phi_{G}^{L,\max}$
but with a size $|G|$ enhancing the absolute effect of $\phi_{G}^{L,\max}$
(more hateful people) instead of having more effect on reducing $\phi_{G'}^{L,\max}$
for all other groups $G^{'}\neq G$ (leader is less credible for everyone
because of high size group). Thus, even when the cost of communication
is $0$, to optimise his returns and maximise the level of hatred
in society, the leader should not target all groups. The relative
sizes of groups and their preferences will be an important factor
in determining whether or not they are being targetted by hateful
messages from the leader to spark hatred.

\subsection{A world with diverse information}

We now focus on a modern world of free diverse information where agents
are not only targetted by signals from the leader but they are also
exposed to a continuum of signals on $(0,1]$.\footnote{We assume that $s_{k}$ can be equal to $1$ exactly since it is the
signal sent from the leader who has some manipulative power whereas
a signal can never be exactly 0.} Since there is a continuum of signals on $(0,1]$ we will simplify
our choice space and assume that the agent has at the begining of
each period $k$ a choice to sample a signal $s_{k}\in\{s^{0},1,\emptyset\}$
where $s^{0}\thickapprox0$ is a signal sent from a policymaker whose
interest is to fight hatred and $s_{k}=1=s^{L}$ is the signal sent
from the leader. Indeed, since the agent choses which signal to sample
based on her realised utility, she is always comparing levels of utility.
We assume that this comparison ends up being a binary comparison between
a signal of almost $0$ and a signal of $1$. \footnote{This is not exactly correct, as signals of different magnitude can
be weighted differently (reflecting accuracy or trust of each source)
and so the comparison is not always transitive. However, in the below,
we will assume that we can always find two signals of very close magnitude
coming from infinite sources which are hence assigned different level
of trust, such that our comparison can be transitive and reduce to
a choice between a signal of magnitude almost $0$ and a signal of
magnitude 1.} If the agent does not sample a signal, she sticks to her prior which
is last period's consensus.

In the presence of multiple signals and a leader targetting at each
period all agents to promote hatred, one might wonder when does the
policymaker have the most return in treating agents with information
to reduce their hate belief. In this set up, agents sample a signal
if and only if it provides the most utility. Thus, signals from the
policymaker might be never sampled and represent a pure economic cost.
In the next proposition, we show when the policymaker has the highest
return in sending truthful information to agents.
\begin{prop}
Suppose the policymaker had the option to randomly send truthful information
($s^{P}\thickapprox0)$ to agents in order to reduce their hate belief.
When agents yield ego utility from their belief and sample a signal
in order to maximise their utility, then the policymaker has the most
return in sending truthful signals when $\sigma$ is distributed on
$\mathcal{U}(0,1)$, where $\mathcal{U}$ is the uniform distribution.
\footnote{E. Weinan, Li, Tiejun and Vanden-Eijnden. 2019. ``Applied Stochastic
Analysis''.\emph{ Graduate Studies In Mathematics. American Mathematical
Society. }has been a useful resource to derive this result.}
\end{prop}
The result of proposition $3$ is very intuititive. The choice of
signal sampled by the agent depends on the value of their $\sigma$.
If the policymaker treats randomly agents with information, she wants
the effect of her treatment to be the greatest. This happens exactly
when $\sigma$ is uniformly distributed. Indeed, if $\sigma$ were
deterministic, the policymaker would be better off learning about
the types of agents beforehands and selecting the agents whose $\sigma$
imply they will sample her signal. Yet, when she treats randomly agents
(because the cost of learning their type is too high for instance)
she maximises the effect of her treatment when $\sigma$ is random
as well. Practically, the policy maker has the most return in sending
informative signals randomly when an event touches a large population
in a similar manner (a pandemic, a large economic crisis...).

Recall that agents weigh the signal from the leader by $\phi_{i}^{L}.$
In the below we will assume that $\lambda_{i}^{P}$, the accuracy
or trust the agent has in the policymaker is $\lambda_{i}^{P}=1-\phi_{i}^{L}.$
Since we restrict signals to be diametrically opposite, they likely
come from antithetic sources. As such, agents will trust one or the
other depending on their identity and preferences. Hence at the beginning
of each period, agents choose to sample or not a signal such that
:

\[
s_{k,i}=\argmax_{\hat{\mu}_{k,i}}U(\hat{\mu}_{k,i})\hspace{1em}\hspace{1em}\hspace{1em}\text{subject to \ensuremath{\hspace{1em}\hspace{1em}}\ensuremath{\hat{\mu}_{k,i}}=\ensuremath{\lambda_{k,i}^{C}}\ensuremath{s_{k,i}^{C}}+(1-\ensuremath{\lambda_{k,i}^{C}})\ensuremath{c_{k-1}}}\hspace{1em}\hspace{1em}\forall k,\hspace{1em}s_{k,i}^{C}\in\{s_{0},1,\emptyset\}
\]

Studying the dynamics of hatred, we get the following result which
describes how society becomes more polarised as the consensus belief
increases. The society becomes split into followers (agents who always
sample the signal of the leader and follow him) and resistants (agents
who follow the policynaker) and the agents who are indifferent decreases.
\begin{prop}
Under the presence of both a leader and a policymaker targetting all
agents with signals respectively $s^{P}\thickapprox0$ and $\mu^{L}=1$,
when agents choose their belief in order to maximise their utility,
then as the consensus group belief increases, the network becomes
more polarised. Eventually, when $c_{k}>\frac{1}{2}$, the society
is completely polarised with agents either following the leader by
sampling his belief or resisting him by sampling the policymaker belief
and these form two distinct groups.
\end{prop}
When $c_{k}<\frac{1}{2}$ , as the consensus increases, more people
are incentivised to sample the signal from the policymaker and similarly
more people are incentivised to sample the signal from the leader.
Equivalently, as the consensus increases, less people have an incentive
to stick to the consensus. This is reflected by different effects
on each group.

Initially, when the consensus is low, a large group of people would
rather stick to the consensus without explicitly taking an individual
risk in either updating their belief towards the policymaker or towards
the leader. Sampling the leader's signal leads to a higher hate belief
and implies higher risk of being punished by acting upon it. On the
reverse, sampling the policymaker signal implies a lower hate belief
and hence imposes a psychological cost on the agent (who can no longer
blame as much on the scapegoat). Therefore, sticking to the consensus
is the conservative option and balances out the two risks. Said differently,
it is less costly for agents to adopt either version of reality passively
rather than explicitly chosing it. However, as the consensus increases,
this group of people becomes smaller as they have more incentive to
adopt either version of reality. The consensus increasing has different
implication on the group of people who are inclined to sample the
signal of the leader vs those who sample the policymaker's.

As the consensus increases, agents with low utility of blaming perceive
a higher risk of being punished by adopting a high hate belief if
they stick to the average group belief and as such make an explicit
decision to chose a lower hate belief by sampling the policymaker's
signal. On the reverse, agents with higher utility of blaming are
more inclined into explicitly adopting a higher hate belief by sampling
the signal of the policymaker since the average group belief is already
high anyways. (i.e. they are incentivised to be more hateful when
other people around them are already somehow hateful).

These two effects are opposite on each group and can be described
of as risk-aversion (or empathy) versus ego utility effect.

Eventually, when $c_{k}>\frac{1}{2}$, the society becomes completely
polarised with agents either explicitly adopting the signal of the
leader or adopting the the signal of the policymaker. On the one hand
agents with high utility of blaming will always follow the leader
(followers) and agents with low utility of blaming (or equivalently
agents who weigh more the practical implications of hatred -lack of
empathy, loss of opportunities in life..) will resist the leader (resistants)
but no one is indifferent.

Given these asymetric dynamics, we can study the equilibrium level
of hatred. In the next proposition, we show that we can characterise
it under different settings.
\begin{prop}
Let $F$ be the function describing the dynamics of the consensus,
i.e. $c_{k}=F(c_{k-1})$ and $c^{*}$ denote its equilibrium if it
exists. We then have the following results:

1. When $\phi^{L}$ and $\sigma$ are independently distributed on
$\mathcal{U}(0,1)$, then $c^{*}=1-\chi$, as $N\to\infty$.

2. Without any assumption on the distribution of $\phi^{L}$ and $\sigma$,
then $F$ is piecewise linear function defined on $[0,1]$ composed
of $M+1$ disjoint intervals $I_{1},...,I_{M}$, where $M$ is the
number of discontinuities of $F$ over $[0,1]$. Then, $F$ admits
a stable fixed point $c^{*}$ if $c^{*}$ is the fixed point of the
first visited interval such that $F(I_{m})\subseteq I_{m}$.

3. When $F$ does not admit a stable fixed point $c^{*}$ in the first
interval where $F(I_{m})\subseteq I_{m}$ and without any assumption
on the distribution of $\phi^{L}$ and $\sigma$, then the consensus
belief is eventually periodic.
\end{prop}
The results of proposition $5$ are striking. Recall that $\sigma$
describes how much an agent weighs the psychological utility yield
from his beliefs relative to its practical utility. As such, $\text{\ensuremath{\sigma} }$
pins down which agents sample the leader or the policymaker's signal
or none. Which agents belong to which category further depends on
the general level of hatred at each period, and hence varies with
time. Consequently, when we do not assume any distribution for $\phi^{L}$
and $\sigma$, the equilibrium level of hatred will depend on the
relative proportions of these three groups, which vary accross periods.
Either these proportions balance out exactly, and $F$ admits a stable
fixed point, or the consensus belief will be eventually periodic.
In such a case, an exact equilibrium is not guaranteed, and the level
of hatred varies periodically. In this equilibrium state, the population
is split into three distinct groups: the resistants, the followers,
and the agents who oscillate between beliefs.Regardless of the bounds
of the periodic oscillations, the value of the punishment $X$ plays
a crucial role in the equilibrium level of hatred. As $X$ increases,
the number of agents who always sample the signal from the policy-maker
(leader) cannot decrease (increase). This suggests that directing
policy towards increasing the practical cost associated with holding
a hate belief has potential to reduce each agent's hatred level.

In fact, in a society where agents' credibility for the leader and
their desire to blame are uniformly distributed, then the equilbrium
level of hatred is solely a function of the punishment associated
with holding a hate belief. This means that if the cause event touched
agents homogeneously, then the best way to reduce the level of hatred,
is to increase the cost associated with it. This observation complements
the findings from Proposition $3$. When $\sigma$ is uniformly distributed,
the policymaker has the highest returns in proposing general information
campaigns, rather than focusing on specific agents in the population.
This is because treating agents randomly, even if they are not the
most hateful, will reduce the general level of hatred via network
feedback effects. In large populations, the equilibrium level of hatred
becomes completely pinned down by the punishment value. This is because
the types of agent as defined by their value of $\{\sigma,\phi^{L}\}$
determine their actions. When $\{\sigma,\phi^{L}\}$ are distributed
uniformly and independently, the asymetric forces guiding agents'
choices eventually balance out in the system. Then, the only threshold
determining the direction of their force (i.e. whether they lean towards
the policymaker or the leader), is only a function of the punishment
$\chi$. Consequently, to reduce the level of hatred, governments
should increase the cost associated with it. This can be understood
either as legal punishment for hate crimes, social cost associated
with being racist...This suggests an important direction for future
policy. 

\section{Conclusion}

Humans optimal decision-making process is often associated with desires
to reduce suffering and maximise happiness. As such, it makes sense
to interpret beliefs within the same framework when they yield utility
to the agent who holds them. We propose a framework to study utilitarian
beliefs within social networks. We believe that it is important to
understand how such beliefs must be analysed differently. They result
from the utility-maximising choice of the agent, where heterogeneity
in preferences implies different individual decision-making processes.
Studying these asymmetric forces within network effects is essential
to better understand how these beliefs evolve. We based our model
on psychological and anthropological evidence for such forces and
applied it to study the dynamics of hatred, -which can be represented
as a utilitarian belief. We find that when preferences are uniformly
distributed, the equilibrium level of hatred depends solely on the
value of the practical punishment associated with holding a hate belief.
Our finding suggests that misinformation campaigns are inefficient
when agents derive utility from their beliefs. An optimal policy should
instead focus on increasing the cost of holding a racist belief. \pagebreak{}

\section*{References}

\leftskip=3em 
\parindent=-3em
\setstretch{1.25}\noindent

Acemoglu, Daron, Victor Chernozhukov, and Muhamet Yildiz. 2016. ``Fragility
of asymptotic agreement under bayesian learning.'' \textit{Theoretical
Economics} 11(1): 187\textendash 225.

Acemoglu, Daron, and Ozdaglar, Asuman. (2010). ``Opinion Dynamics
and Learning in Social Networks''. Working Paper

Acemoglu, Daron, Dahleh, M.A., Lobel, Ilan, and Ozdaglar, Asuman.
(2008). ``Bayesian Learning in Social Networks''. \emph{Review of
Economic Studies. 78:1201-1236.}

Argyle, M. (1987). ``The psychology of happiness''. Methuen.

Banerjee, A. (1992). ``A Simple Model of Herd Behavior''. \emph{The
Quarterly Journal of Economics}. 107(3): 797\textendash 817

Banerjee, Abhijit, and Fudenberg, Drew. (2004). ``Word-of-Mouth Learning''.
\emph{Games and Economic Behavio}r. 46(1):1-22

Baumeister, R. F., and Leary, M. R. (1995). ``The need to belong:
Desire for interpersonal attachments as a fundamental human motivation''.
\emph{Psychological Bulletin}, 117(3), 497\textendash 529.

Baumeister, R. F., and Wotman, S. R. (1992).'' Emotions and social
behavior.Breaking hearts: The two sides of unrequited love''. Guilford
Press.

Baumeister, R. F., and Tice, D. M. (1990). ``Anxiety and social exclusion''.
\emph{Journal of Social and Clinical Psychology}: 9(2), 165\textendash 195.

Beardsley, Elizabeth. (1970). ``Moral Disapproval and Moral Indignation''.
Philosophy and Phenomenological Research (31): 161-176

Bikhchandani, Sushil, David Hirshleifer, and Ivo Welch. (1992).``A
Theory of Fads, Fashion, Custom, and Cultural Change as Informational
Cascades''. \emph{Journal of Political Economy}. 100(5): 992-1026.

Caplin, Andrew, and Leahy, John. (2001). ``Psychological Expected
Utility Theory and Anticipatory Feelings''. \emph{The Quarterly Journal
of Economics. 116(1):55-79}

Chen, Yan, and Sherry Xin Li. 2009. \textquotedbl Group Identity
and Social Preferences.\textquotedbl{} American Economic Review, 99
(1): 431-57

Cikara M, Botvinick MM and Fiske ST. (2011). ``Us versus them: social
identity shapes neural responses to intergroup competition and harm''.\emph{
Psychol Sci}. 22(3): 306-13

Degroot, Morris H. 1974. ``Reaching a Consensus''. \emph{Journal
of the American Statistical Association. }69(345) : 118-121.

De Marzo, Peter, Vayanos, Dimitri, and Zwiebel, Jeffrey. (2003). ``Persuasion
Bias, Social Influence, and Unidimensional Opinions''. \emph{The
Quarterly Journal of Economics}. (118)-3: 909\textendash 968.

Dollard, J., Miller, N. E., Doob, L. W., Mowrer, O. H., \& Sears,
R. R. (1939). Frustration and aggression. Yale University Press.

Douglas, T. (1995). Scapegoats: Transferring blame. New York, NY:
Routledge Press.

Ellison, Glenn, and Fudenberg, Drew. (1993). ``Rules of thumb for
social learning''. \emph{Journal of Political Economy} . 101(4):
612-643

E. Weinan, Li, Tiejun and Vanden-Eijnden. 2019. ``Applied Stochastic
Analysis''.\emph{ Graduate Studies In Mathematics. American Mathematical
Society.}

Festinger, L. (1950). ``Informal social communication''. \emph{Psychological
Review}. 57(5): 271\textendash 282

Festinger, L., Schachter, S., and Back, K. (1950). ``Social pressures
in informal groups; a study of human factors in housing''. Harper

Freud, Sigmund. 1896. ``Further Remarks on the Neuro-Psychoses Of
Defence - Weitere Bemerkungen Über Die Abwehrneuropsychosen'' Neurol.
Zbl., 15(10): 434-48.

Freud, Sigmund. 1905. ``Fragment of An Analysis Of A Case Of Hysteria
- Bruchstück einer Hysterie-Analyse''. Mschr Psychiat Neurol (18):285\textendash 309.

Galofré-Vilà, Gregori, Meissner, Christoper M., McKee, Martin and
Stuckler, David. 2017 ``Austerity and the rise of the Nazi party''.
\emph{NBER Working Paper 24106.}

Girard, René. 1982. ``Le bouc émissaire''. \emph{Editions Grasset}.

Glaeser, Edward L. 2005. ``The Political Economy of Hatred''. \emph{The
Quarterly Journal of Economics. }120(1):45-86.

Glick, P. (2002). ``Sacrificial lambs dressed in wolves' clothing:
Envious prejudice, ideology, and the scapegoating of Jews.'' In \emph{L.
S. Newman \& R. Erber (Eds.), Understanding genocide: The social psychology
of the Holocaust }(p. 113\textendash 142). Oxford University Press.

Glick, P. (2005). Choice of Scapegoats. In \emph{J. F. Dovidio, P.
Glick, \& L. A. Rudman (Eds.), On the nature of prejudice: Fifty years
after Allport }(p. 244\textendash 261). Blackwell Publishing.

Golub, Benjamin, and Jackson, Matthew O. (2010). ``Naïve Learning
in Social Networks and the Wisdom of Crowds''. American Economic
Journal: Microeconomics, 2 (1): 112-49

Gurdal, M.Y., Miller J.B. and Rustichini, A. (2013). ``Why Blame?''.
\emph{Journal of Political Economy}. 121(6):1205-1246. The University
of Chicago Press

Hovland, C. I., \& Sears, R. R. (1940). Minor studies of aggression:
VI. Correlation of lynchings with economic indices. The Journal of
Psychology: Interdisciplinary and Applied, 9, 301\textendash 310.

Hilton, D. J. (2007). ``Causal explanation: From social perception
to knowledge-based causal attribution''. In \emph{A. W. Kruglanski
\& E. T. Higgins (Eds.), Social psychology: Handbook of basic principles
(2nd ed., pp. 232\textendash 253)}. New York, NY: Guilford

Kemeny, J.G. and Snell, J.L. (1960). ``Finite Markov Chains''.

Kiecolt-Glaser, J.K. et al. (1984). ``Psychosocial modifiers of immunocompetence
in medical students''. \emph{Psychosom Med}: 46(1):7-14

Köszegi, Botond. (2003). ``Ego Utility, Overconfidence, and Task
Choice''. \emph{Journal of the European Economic Association. }4(4):673-707

Leary, M. R., Tambor, E. S., Terdal, S. K., \& Downs, D. L. (1995).
``Self-esteem as an interpersonal monitor: The sociometer hypothesis''.
\emph{Journal of Personality and Social Psychology}: 68(3), 518\textendash 530

Lynch, J. J. (1977). ``The broken heart: the medical consequences
of loneliness in America''. New York: Basic Books.

Malle, B. F., Guglielmo, S., \& Monroe, A. E. (2014). ``A theory
of blame''. \emph{Psychological Inquiry}. 25(2): 147\textendash 186.

Malle, B. F. and Knobe, J. (1997). ``The folk concept of intentionality''.
\emph{Journal of Experimental Social Psychology.} 33(2): 101\textendash 121

Mangel, M, and Clark, C.W. (1986). ``Towards a Unified Foraging Theory''.
\emph{Ecology, 67(5):1127-1138.}

McKenna, M. (2012). Directed blame and conversation. In \emph{Blame:
Its nature and norms:} 119\textendash 140. New York, NY: Oxford University
Press.

Moise, Jean. 2014. ``The Rwandan Genocide: The True Motivations for
Mass Killings''.

Molavie, Pooya, Tahbaza-Salehi, Alireza, and Jadbabaie Ali. (2018).
Econometrica. 86(2):445-490.

Mossel, Elchanan, Sly, Allan, and Tamuz, Omer. (2015). ``Strategic
Learning and the Topology of Social Networks''. \emph{Econometrica}.
83(5):1755-1794.

Myers, D. (1992).The pursuit of happiness.New York: Morrow

Nogueira, A. and Pires, B., Rosales, R. A., (2013) Asymptotically
periodic piecewise contractions of the interval Nonlinearity, IOP
Publishing, 2014, 27 (7), pp.1603-{}-1610.

Nogueira, A. and Pires, B., (2012). Dynamics of piecewise contractions
of the interval. Ergodic Theory and Dynamical Systems, Cambridge University
Press (CUP), 2015, 35 (07), pp.2198-2215.

Oster, Emily, Ira Shoulson, and E. Ray Dorsey. 2013. \textquotedbl Optimal
Expectations and Limited Medical Testing: Evidence from Huntington
Disease.\textquotedbl\emph{ American Economic Review}. 103 (2):
804-30.

Rothschild ZK, Landau MJ, Sullivan D, Keefer LA. 2012. ``A dual-motive
model of scapegoating: displacing blame to reduce guilt or increase
control''. \emph{J Pers Soc Psychol}. 102(6):1148-63.

Schachter, S. (1951). ``Deviation, rejection, and communication''.
\emph{The Journal of Abnormal and Social Psychology}: 46(2), 190\textendash 207

Tajfel, H, Billig, M.G., Bundy, R.P. and Flament, Claude. (1971).
``Social categorization and intergroup behaviour''. \emph{European
Journal of Social Psychology 1(2):149-178.}

Tajfel, H. (1981). ``Human Groups and Social Categories: Studies
in Social Psychology''\emph{. }Cambridge University Press.

Tiedens, L. Z. and C. W. Leach (Eds.). (2004). ``Studies in emotion
and social interaction. The social life of emotions'' (p. 314\textendash 335

Yariv, Leeat. (2001). ``Believe and Let Believe: Axiomatic Foundations
for Belief Dependent Utility Functionals''. Available at SSRN.

Yariv, Leeat. (2002). ``I'll See it When I Believe it ? A Simple
Model of Cognitive Consistency''. Available at SSRN.

Wong, P. T., and Weiner, B. (1981). ``When people ask \textquotedbl why\textquotedbl{}
questions, and the heuristics of attributional search''. \emph{Journal
of Personality and Social Psychology}. 40(4): 650\textendash 663''

\pagebreak{}

\appendix

\section{Appendix : Proofs}

\subsection*{Proof of Proposition 1}
\begin{proof}
First, let's prove the finite horizon case.

In period $0$, the consensus is $c_{0}=\omega_{i}^{*}\widehat{\mu}_{0,i}$,
where $\hat{\mu}_{0}$ is either $0$ or $\phi_{i}^{L}$. Agents start
period $0$ with no hatred, i.e. $\mu_{-1,i}=0$ for all $i$ and
are xposed to $s_{0}=\mu_{L}=1$. They can choose to adopt it weighted
by $\phi_{i}^{L}$, where $\phi_{i}^{L}>0$ in which case $\widehat{\mu}_{0,i}=\phi_{i}^{L}>0$
or stick to his prior in which case $\hat{\mu}_{0,i}=$$\mu_{-1,i}=0$.

\textbf{Utility if she adopts the leader belief}

\begin{align*}
\mathbb{E}(U|x_{0}=\mu_{L}) & =\phi_{i}^{L}(-X)+(1-\phi_{i}^{L})X+\sigma_{i}\phi_{i}^{L}\\
 & =-2X\phi_{i}^{L}+X+\sigma_{i}\phi_{i}^{L}
\end{align*}

\textbf{Utility if she sticks to his prior}

\begin{align*}
\mathbb{E}(U|x_{0}=\emptyset) & =\mu_{-1}(-X)+(1-\mu_{-1})X\\
 & =X
\end{align*}

Thus an agent adopts the leader belief in period 0 if and only if
$\mathbb{E}(U|x_{0}=\mu_{L})>\mathbb{E}(U|x_{0}=\emptyset)$ which
happens when $\sigma_{i}>2X$

Consequently after their utility maximisation choice, agents have
either $\widehat{\mu}_{0,i}=\phi_{i}^{L}$ if $\sigma_{i}>2X$ or
$\widehat{\mu}_{0,i}=\mu_{-1}=0$ if $\sigma_{i}\leq2X$ .

Then $c_{0}=\sum_{i|\sigma_{i}>2X}\omega_{i}^{*}\phi_{i}^{L}+\sum_{i|\sigma_{i}\leq2X}\omega_{i}^{*}\cdot0=\sum_{i|\sigma_{i}>2X}\omega_{i}^{*}\phi_{i}^{L}$
. It follows that $\min_{\omega^{*}}c_{0}\iff\omega_{j}^{*}\leq\omega_{k}^{*}\text{ for every}\sigma_{j}\geq\sigma_{k}\text{ and }\phi_{j}^{L}<\phi_{k}^{L}.$

Let's now prove the infinite case now.

Let's evaluate the utility at any period $k>0.$ In any other period,
the agent is evaluating the utility if he adopts the leader belief

\begin{align*}
\mathbb{E}(U|x_{1}=\mu_{L}) & =\phi_{i}^{L}(-X)+(1-\phi_{i}^{L})X+\sigma_{i}\phi_{i}^{L}\\
 & =-2X\phi_{i}^{L}+X+\sigma_{i}\phi_{i}^{L}
\end{align*}

vs the utility if he sticks to his prior:

\begin{align*}
\mathbb{E}(U|x_{2}=\emptyset) & =c_{k-1}(-X)+(1-c_{k-1})X\\
 & =-2c_{k-1}X+X
\end{align*}

Indeed,at round $k$, an agent adopts the leader belief if $\phi_{i}^{L}>\mu_{k-1}$
and $\sigma_{i}>\frac{2X(\phi_{i}^{L}-c_{k-1})}{\phi_{i}^{L}}$or
sticks to last period consensus belief $c_{k-1}$.

We can rewrite the consensus in period $k$ as :

\[
c_{k}=\sum_{\left\{ i:\sigma_{i}>\frac{2X(\phi_{i}^{L}-c_{k-1})}{\phi_{i}^{L}},\phi_{i}^{L}>c_{k-1}\right\} }w_{i}\phi_{i}^{L}+\sum_{\left\{ i:\sigma_{i}<\frac{2X(\phi_{i}^{L}-c_{k-1})}{\phi_{i}^{L}}\right\} }w_{i}c_{k-1}+\sum_{\left\{ i:\sigma_{i}>\frac{2X(\phi_{i}^{L}-c_{k-1})}{\phi_{i}^{L}},\phi_{i}^{L}<c_{k-1}\right\} }w_{i}c_{k-1}
\]

\[
c_{k}=\sum_{\left\{ i:\sigma_{i}>\frac{2X(\phi_{i}^{L}-c_{k-1})}{\phi_{i}^{L}},\phi_{i}^{L}>c_{k-1}\right\} }w_{i}\phi_{i}^{L}+c_{k-1}\left(\sum_{\left\{ i:\sigma_{i}<\frac{2X(\phi_{i}^{L}-c_{k-1})}{\phi_{i}^{L}}\right\} }w_{i}+\sum_{\left\{ i:\sigma_{i}>\frac{2X(\phi_{i}^{L}-c_{k-1})}{\phi_{i}^{L}},\phi_{i}^{L}<c_{k-1}\right\} }w_{i}\right)
\]

Let's prove the first result. First note that as long as there is
an agent for whom $\sigma_{i}>\frac{2X(\phi_{i}^{L}-c_{k-1})}{\phi_{i}^{L}},\phi_{i}^{L}>c_{k-1}$
then $c_{k}>c_{k-1}$ .

If there is at least one agent $j$ with $\sigma_{j}>2X$, he adopts
$\phi_{j}^{L}$ in period 0, then $c_{0}>0$ and $c_{k}$ is updated
dynamically as more agents have an incentive to switch to hatred.
Indeed, in period $k$, the incentive to switch to hatred is $\sigma_{i}>\frac{2X(\phi_{i}^{L}-c_{k-1})}{\phi_{i}^{L}}$
whereas, in period $k+1$ it is $\sigma_{i}>\frac{2X(\phi_{i}^{L}-c_{k})}{\phi_{i}^{L}}$
.

Let the consensus becomes stable from period $K^{*}$ onwards. Then,
in period $K^{*}$ all agents have $\mu_{k^{*}}=c.$ Assume for the
sake of contradiction that $c<\max\{\phi_{i}^{L}\}_{i:\sigma_{i}>\frac{2X(\phi_{i}^{L}-c_{q})}{\phi_{i}^{L}},\text{for some q}}$.
Let agent $m$ be the agent such that $\phi_{m}^{L}=\max\{\phi_{i}^{L}\}_{i:\sigma_{i}>\frac{2X(\phi_{i}^{L}-c_{q})}{\phi_{i}^{L}},\text{for some q}}$
but $\phi_{m}^{L}>c$.

Then in period $K^{*},$ agent $m$ choose $\phi_{m}^{L}$ which contradicts
our initial assumption.

Thus, $lim_{k\to\infty}c_{k}=\max\{\phi_{i}^{L}\}_{i:\sigma_{i}>\frac{2X(\phi_{i}^{L}-c_{q})}{\phi_{i}^{L}},\text{for some q}}$.

To prove the first result, let $j$ be an agent such that $\sigma_{i}>2X$
and $\phi_{j}^{L}=\max\{\phi_{i}^{L}\}_{i\in N}$.

Then by the above, if agent $j$ has incentive to switch in period
$0$ to $\phi_{j}^{L}$, then at each period he chooses $\phi_{j}^{L}$.
By definition, at stationary state, $\mu_{i}=c\hspace{1em}\forall i$
.

Again by contradiction, at stationary state, agent $c=\phi_{j}^{L}$.

Thus $\lim_{k\to\infty}c_{k}=\max\{\phi_{i}^{L}\}_{i\in N}$
\end{proof}

\subsection*{Proof of Proposition 2}

Let $S$ be the set of groups from which the leader can choose. $S=\{G_{1},G_{2},...,G_{k}\}$.

We define his return $R:2^{S}\to\mathbb{R}$ and for a given set $S^{'}\subseteq S$,
we have :

$R(S^{'})=\sum_{g\in S^{'}}(c_{r,g}|g|-(r+1)c|g|)$

Assuming $c=0$,

$R(S^{'})=\sum_{g\in S^{'}}c_{r,g}|g|$

Since $r$is a period of equilibrium, then we have showed above that
$c_{r,g}=\max_{i\in g}\{\phi_{i}^{L}\}_{i:\sigma_{i}>\frac{2X(\phi_{i}^{L}-c_{q})}{\phi_{i}^{L}},\text{for some q}}$.
In large groups, we can assume that the condition $i:\sigma_{i}>\frac{2X(\phi_{i}^{L}-c_{q})}{\phi_{i}^{L}},\text{for some q}$
is satisfied, and hence $c_{r,g}=\max_{i\in g}\phi_{i}^{L}=\phi_{g}$.

Then we defined $\phi_{i,S}^{L}=\max(0,\phi_{i}^{L}-\frac{|\cup_{g\in S}g|}{K})$,
so $\phi_{g,S}^{L}=\max(0,\phi_{g}^{L}-\frac{|\cup_{g\in S}g|}{K})$.

Since we already defined $\phi_{G}=\max_{i\in g}\phi_{i}^{L}$, then
we will let $\phi_{g,S}^{L}>0,$ hence $\phi_{g,S}^{L}=\phi_{g}^{L}-\frac{|\cup_{g\in S}g|}{K}.$

Then 
\begin{align*}
R(S^{'})= & \sum_{g\in S^{'}}(\phi_{g}-\frac{|\cup_{g\in S^{'}}g|}{K})|g|\\
= & \sum_{g\in S^{'}}(\phi_{g}|g|-\frac{|\cup_{g\in S^{'}}g|\cdot|g|}{K})\\
\end{align*}

Since we defined social groups such that $g_{i}\cap g_{j}=\emptyset\hspace{1em}\forall i\neq j$
then,

\begin{align*}
R(S^{'})= & \sum_{g\in S^{'}}(\phi_{g}|g|-\frac{|g|^{2}}{K})
\end{align*}

In the case with homogenous preferences, $\phi_{G}=\phi\hspace{1em}\forall g$

\begin{align*}
R(S^{'})= & \sum_{g\in S^{'}}(\phi\cdot|g|-\frac{|g|^{2}}{K})
\end{align*}

Then the return becomes only a function of the total size of the groups
chosen in a given set $S^{'}$. Let $N=\sum_{g\in S^{'}}|g|$ for
a given set $S^{'}$.

We can rewrite 
\begin{align*}
R(N)= & \phi N-\frac{N^{2}}{K}
\end{align*}

Then we see that the return maximising set $S^{'}$ is characterised
by $N=\phi\frac{K}{2}$

In the case with non-homegenous preferences but equal size groups,
we can rewrite

\begin{align*}
R(S^{'})= & \sum_{g\in S^{'}}(\phi_{G}\cdot M\cdot N-\frac{M^{2}N^{2}}{K})\\
= & MN\sum_{g\in S^{'}}(\phi_{G}-\frac{MN}{K})
\end{align*}

where $M=|S^{'}|$ and $N=|g|,\hspace{1em}\forall g$

Then notice that adding a group increases return by $\phi_{G}-\frac{N}{K}$.
Consequently, the leader will target all groups $g$ with $\phi_{g}>\frac{N}{K}$.

Finally, under non-homegenous preferences and unequal group sizes,
then, we can rewrite

\begin{align*}
R(S^{'})= & \sum_{g\in S^{'}}(\phi_{g}|g|-\frac{|g|^{2}}{K})\\
= & \sum_{g\in S^{'}}\phi_{g}|g|-\sum_{g\in S^{'}}\frac{|g^{2}|}{K}-2\frac{\sum_{j=1}^{M}\sum_{i=1}^{j-1}|g_{i}|\cdot|g_{j}|}{K}
\end{align*}

If $S^{'}$ is a return-maximising set, then if $g_{m}$ is an omitted
group in equilibrium, we must have

$R(S^{'}\cup g_{m})-R(S^{'})<0\iff\phi_{g_{m}}|g_{m}|-\frac{|g_{m}|^{2}}{K}-2\sum_{l\in S^{'}}\frac{|g_{m}|\cdot|g_{l}|}{K}<0$

Thus all omited groups $g_{m}\notin S^{'}$ are such that $\phi_{g_{m}}<\frac{|g_{m}|}{K}+2\sum_{l\in S^{'}}\frac{|g_{l}|}{K}$.

And on the contrary, all included groups, $g_{s}\in S^{'}$ are such
that $\phi_{g_{s}}<\frac{|g_{s}|}{K}+2\sum_{l\in S^{'},l\neq s}\frac{|g_{l}|}{K}$

\subsection*{Proof of Proposition 3}

Define the outcome $x_{j}$ to be the event that when the policy maker
sends a signal $x_{k}=0$ to agent $j$, the consensus in our network
$c_{k}$ ends up being lower versus when they do not.

We have,

\begin{align*}
p_{j}= & \mathbb{P}(X=x_{j})\\
= & \mathbb{P}(c_{k}|\text{ agent \ensuremath{j} gets signal \ensuremath{x_{k}=0} }<c_{k}|\text{ agent \ensuremath{j} does not get signal }x_{k}=0)\\
= & \mathbb{P}(\text{agent j samples signal \ensuremath{x_{k}=0}})\\
= & \mathbb{P}(U_{k,j}(\text{agent j samples signal \ensuremath{x_{k}=0}})>U_{k,j}(\text{agent j does not sample signal \ensuremath{x_{k}=0}}))\\
= & \begin{cases}
\mathbb{P}(\sigma_{j}\leq2Xc_{k-1}) & \mbox{if }c_{k-1}\leq\frac{1}{2}\\
\mathbb{P}\left(\sigma_{j}<2X(\phi_{j}^{L}+c_{k-1}-2\phi_{j}^{L}c_{k-1})\right) & \mbox{if }c_{k-1}>\frac{1}{2}
\end{cases}
\end{align*}

We can then define our Shannon entropy to provide the expected information
content of $X$:

\[
H(X)=\mathbb{E}(I(X))=\mathbb{E}(-log(\mathbb{P}(x_{j}))
\]

We then get:

\[
H(X)=-\sum_{j=1}^{n}p_{j}log(p_{j})
\]

Let's prove that $H$ is maximized when $X$ and hence $\sigma$ is
uniformely distributed.

We define the function $f$ on the range (0, 1{]} as follows:

\[
f(x)=x\log(x)
\]

We can then write:

\[
H(X)=-\sum_{j=1}^{n}f(p_{j})
\]

The second derivative of $f$ is:

\[
f^{''}(x)=\frac{1}{x}
\]

which is strictly positive on the (0, 1{]}. Hence $f$ is strictly
convex on (0, 1{]}, and the sum $\sum_{j=1}^{n}f(p_{j})$ is a stricly
convex function. Thus, $H$ is strictly concave. From the convexity
of $f$, we make use of the Jensen's inequality applied to the random
variable $\mathbb{P}(X)$, to write the following:

\begin{align*}
f\left(\mathbb{E}(\mathbb{P}(X))\right)\leq\mathbb{E}\left(f(\mathbb{P}(X))\right)\implies & f\left(\frac{\sum_{j=1}^{n}p_{j}}{n}\right)\leq\frac{1}{n}\sum_{j=1}^{n}f(p_{j})\\
\implies & \frac{1}{n}\sum_{j=1}^{n}f(p_{j})\geq f\left(\frac{1}{n}\right)\\
\implies & -H(X)\geq n\frac{1}{n}\log\left(\frac{1}{n}\right)\\
\implies & H(X)\leq\log(n)
\end{align*}

For a uniform distribution, where $p_{j}=\frac{1}{n},\forall j$,
the Shannon entropy is:

\[
H(X)=-\sum_{j=1}^{n}p_{j}log(p_{j})=-n\frac{1}{n}\log\left(\frac{1}{n}\right)=\log(n)
\]

which achieves the upper bound limit.

Finally, as $H$ is a striclty concave function, the uniform distribution
is the unique distribution which maximizes the Shannon entropy.

Intuitively this mean that information has highest power when $\sigma$
is uniformely distributed. Since a signal $x=0$ can only reduce the
consensus, then the effect of information is strictlly positive. Therefore,
maximising the expected information content of our information release
is equivalent to maximising the return of information.

\subsection*{Proof of Proposition 4}

Fix a period $k$ and remember that now the agent has a choice to
sample three signals $s_{k}\in\{s^{P},s^{L},\emptyset\}$ where $s^{P}\thickapprox0$
is the signal from the policymaker, $s^{L}=\mu^{L}=1$ is the signal
from the leader. Let's evaluate the utility under each sampling possibility.
\begin{itemize}
\item \textbf{Utility if the agent samples the signal from the policy-maker}
\end{itemize}
The belief of the agent becomes $\hat{\mu}_{k,i}=\lambda_{i}^{P}s^{P}+(1-\lambda_{i}^{P})c_{k-1}\thickapprox(1-\lambda_{i}^{P})c_{k-1}$
since $s_{k}\thickapprox0$ and the expected utility $U$ is:
\begin{align*}
E(U(\hat{\mu}_{k,i}|s=s_{k}^{P}=0))= & X(1-(1-\lambda_{i}^{P})c_{k-1})-X(1-\lambda_{i}^{P})c_{k-1}-\sigma_{i}\lambda_{i}^{P}\\
= & X-2X(1-\lambda_{i}^{P})c_{k-1}-\sigma_{i}\lambda_{i}^{P}
\end{align*}

\begin{itemize}
\item \textbf{Utility if the agent samples the signal from the leader}
\end{itemize}
The belief of the agent becomes $\hat{\mu}_{k,i}=\phi_{i}^{L}+(1-\phi_{i}^{L})c_{k-1}$
and the expected utility $U$ is:
\[
\begin{aligned}E(U(\hat{\mu}_{k,i}|s=s_{k}^{L}=1))= & X(1-(\phi_{i}^{L}+(1-\phi_{i}^{L})c_{k-1}))-X(\phi_{i}^{L}+(1-\phi_{i}^{L})c_{k-1})+\sigma_{i}\phi_{i}^{L}\\
= & X-2X\phi_{i}^{L}-2X(1-\phi_{i}^{L})c_{k-1}+\sigma_{i}\phi_{i}^{L}
\end{aligned}
\]

\begin{itemize}
\item \textbf{Utility if the agent sticks to its prior}
\end{itemize}
The belief of the agent becomes $\hat{\mu}_{k,i}=c_{k-1}$ and the
expected utility $U$ is:
\begin{align*}
E(U(\hat{\mu}_{k,i}|s_{k}=\emptyset))= & X(1-c_{k-1})-Xc_{k-1}\\
= & X-2Xc_{k-1}
\end{align*}

The agent will sample from the policy-maker or the leader iff it provides
the highest utility of the three methods, assume that $\lambda_{i}^{P}=1-\phi_{i}^{L}$.

Then the agent prefers sampling the signal of the leader over sticking
to the consensus when

\textbf{
\begin{align*}
E(U(\hat{\mu}_{k,i}|s=s_{k}=1))> & E(U(\hat{\mu}_{k,i}|s=s_{k}=\emptyset))\\
\iff X-2X\phi_{i}^{L}-2X(1-\phi_{i}^{L})c_{k-1}+\sigma_{i}\phi_{i}^{L}> & X-2Xc_{k-1}\\
\iff\sigma_{i}> & 2X(1-c_{k-1})
\end{align*}
}

Then the agent prefers sampling the signal of the leader over the
policymaker when

\textbf{
\begin{align*}
E(U(\hat{\mu}_{k,i}|s=s_{k}=1))> & E(U(\hat{\mu}_{k,i}|s=s_{k}=0))\\
\iff\sigma_{i}> & \frac{2X(\phi_{i}^{L}+c_{k-1}(1-2\phi_{i}^{L}))}{1}\\
\sigma_{i}> & 2X(c_{k-1}+\phi_{i}^{L}-2\phi_{i}^{L}c_{k-1})
\end{align*}
}

Finally, the agent prefers sticking to the consensus over sampling
the policymaker when:

\textbf{
\begin{align*}
E(U(\hat{\mu}_{k,i}|s=s_{k}=\emptyset))> & E(U(\hat{\mu}_{k,i}|s=0))\\
\iff X-2Xc_{k-1}> & X-2X(1-\lambda_{i}^{P})c_{k-1}-\sigma_{i}\lambda_{i}^{P}\\
\iff\sigma_{i}> & 2Xc_{k-1}
\end{align*}
}

We can then define :

\begin{align*}
\alpha & =2X(1-c_{k-1})\\
\beta & =2X(c_{k-1}+\phi_{i}^{L}-2\phi_{i}^{L}c_{k-1})\\
\gamma & =2Xc_{k-1}
\end{align*}

We will then rank $\alpha,\beta,\lambda$ on $[0,1]$
\begin{enumerate}
\item Rank $\alpha,\beta$
\[
\begin{aligned}\alpha<\beta\iff & 1-c_{k-1}<c_{k-1}+\phi_{i}^{L}-2\phi_{i}^{L}c_{k-1}\\
\iff & 1-2c_{k-1}<\phi_{i}^{L}(1-2c_{k-1})\\
\iff & \phi_{i}^{L}>1,c_{k-1}<\frac{1}{2}\quad\mbox{or}\quad\phi_{i}^{L}<1,c_{k-1}>\frac{1}{2}\\
\iff & c_{k-1}>\frac{1}{2}\quad\mbox{as}\quad\phi_{i}^{L}<1\forall i.
\end{aligned}
\]
\item Rank $\text{\ensuremath{\alpha,\gamma}}$

\[
\alpha<\gamma\iff1-c_{k-1}<c_{k-1}\iff c_{k-1}>\frac{1}{2}
\]

\item Rank $\beta,\gamma$

\[
\beta<\gamma\iff c_{k-1}+\phi_{i}^{L}-2\phi_{i}^{L}c_{k-1}<c_{k-1}\iff c_{k-1}>\frac{1}{2}
\]

\end{enumerate}
We then have two cases

\textbf{Case 1 }$c_{k-1}<\frac{1}{2}$

Then we have $\begin{cases}
\alpha>\beta\\
\alpha>\gamma\\
\beta>\gamma
\end{cases}$

Hence we have $\alpha>\beta>\gamma$

Let $\sigma_{i}\in[0,1]$ where $0\leq\gamma<\beta<\alpha\leq1$.
We evaluate the preferences on each interval. Define $\mathcal{C=}\{L,P,C\}$
to be the set of choices for the agent where he can of sampling the
leader signal, the policmaker's and sticking to the cosnensus. Per
above each is associate with an expected utility level so we can establish
preferences.

In $[0,\gamma]$, we have $C\gtrsim L,P\gtrsim L$ and $P\gtrsim C$,
hence by transitivity, the agent prefers to sample the policymaker's
signal.

In $(\gamma,\beta]$, we have $C\gtrsim L,P\gtrsim L$ and $C\gtrsim P$,
hence by transitivity, the agent prefers to stick to the consensus.

In $(\beta,\alpha]$, we have $C\gtrsim L,L\gtrsim P$ and $C\gtrsim P$,
hence by transitivity, the agent prefers to stick to the consensus.

In $(\alpha,1]$, we have $L\gtrsim C,L\gtrsim P$ and $C\gtrsim P$,
hence by transitivity, the agent prefers to sample the leader.'s signal.

Therefore we get a complete mapping on $[0,1]$ for the choices of
sampling by the agent :

$\begin{cases}
\text{Agent samples the policymaker signal} & \iff\sigma_{i}\in[0,2Xc_{k-1}]\\
\text{Agent sticks to the consensus} & \iff\sigma_{i}\in(2Xc_{k-1},2X(1-c_{k-1})]\\
\text{Agent samples the leader signal } & \iff\sigma_{i}\in(2X(1-c_{k-1}),1]
\end{cases}$

Consequently we see that as $c_{k-1}\uparrow$, less people choose
to stick to the consensus.

\vspace{10pt}

\textbf{Case 2 }$c_{k-1}>\frac{1}{2}$

Then we have $\begin{cases}
\alpha<\beta\\
\alpha<\gamma\\
\beta<\gamma
\end{cases}$

Hence we have $\alpha<\beta<\gamma$

In $[0,\alpha]$, we have $C\gtrsim L,P\gtrsim L$ and $P\gtrsim C$,
hence by transitivity, the agent prefers to sample the policymaker's
signal.

In $(\alpha,\beta]$, we have $L\gtrsim C,P\gtrsim L$ and $P\gtrsim C$,
hence by transitivity, the agent prefers to sample the policymaker's
signal.

In $(\beta,\gamma]$, we have $L\gtrsim C,L\gtrsim P$ and $P\gtrsim C$,
hence by transitivity, the agent prefers to sample the leader's signal.

In $(\gamma,1]$, we have $L\gtrsim C,L\gtrsim P$ and $C\gtrsim P$,
hence by transitivity, the agent prefers to sample the leader's signal.

Therefore we get a complete mapping on $[0,1]$ for the choices of
sampling by the agent :

$\begin{cases}
\text{Agent samples the policymaker signal} & \iff\sigma_{i}\in[0,2X(c_{k-1}+\phi_{i}^{L}-2\phi_{i}^{L}c_{k-1})]\\
\text{Agent sticks to the leader signal } & \iff\sigma_{i}\in(2X(c_{k-1}+\phi_{i}^{L}-2\phi_{i}^{L}c_{k-1}),1]
\end{cases}$

Consequently when $c_{k-1}>\frac{1}{2}$, the choice space becomes
completely polarised, and agents adopt either the policymaker's signal
or the leader signal.

\subsection*{Proof of Proposition 5}

\textbf{Part 1. }\emph{When the punishment value $X$ is such that
$X\geq\frac{1}{2}$, agents never sample the leader's signal and $\forall k\in\mathbb{N}$
the consensus belief $c_{k}=0$.}
\begin{proof}
Let's prove that if $X\geq\frac{1}{2}$, then $\forall k\in\mathbb{N}$
the consensus belief remains $c_{k}=0$.

We will prove the proposition by induction.

Let $X\geq\frac{1}{2}$ and suppose $c_{k-1}=0$, $k-1\in\mathbb{N}$.
We will show that no agent will sample from the leader nor the policy-maker.
Thus, the consensus belief will remain $c_{k}=0$. 

From the Proof of Proposition 4, with $c_{k-1}=0$, we write that
agents will sample the leader's signal:
\[
\iff\sigma_{i}>2X(1-c_{k-1})=2X\geq1.
\]

We also note that agents will sample the policy-makers's signal:
\[
\iff\sigma_{i}\leq2Xc_{k-1}=0.
\]

Recall that:
\[
\forall i\in n,\sigma_{i}\in(0,1).
\]

Thus, no agent sample from the leader nor the policy-maker at timepoint
$k$. Each agent will stick to the consensus and hold a belief $\mu_{i,k}=c_{k-1}=0$.

Therefore,
\[
c_{k}=\sum_{i=1}^{n}w_{i}\mu_{i,k}=0.
\]

We proved that:
\[
\forall k\in\mathbb{N},X\geq\frac{1}{2},c_{k-1}=0\implies c_{k}=0.
\]

We note that the consensus belief at timepoint $k=0$ is such that
$c_{k}=0$. Thus $c_{1}=0$ and we proved by induction that:
\[
X\geq\frac{1}{2}\implies\forall k\in\mathbb{N},c_{k}=0.
\]
\end{proof}
\uline{\pagebreak}\textbf{Part 2.} \emph{For $X<\frac{1}{2}$,
if we assume that $\sigma_{i}$ and $\phi_{i}^{L}$ are realizations
of i.i.d uniformly distributed random variables $\Sigma$ and $\Phi^{L}$
over $(0,1)$ respectively, and let the number of agents $n$ tend
to infinity, then the sequence of consensus belief $c_{k}=f(c_{k-1})\forall k\in\mathbb{N^{*}}$
starting at $c_{0}=0$ is monotonically increasing and converges to
$c_{\infty}=1-X$.}
\begin{proof}
We will follow three steps to prove Proposition 2: 1) we will show
that by assuming $\sigma_{i}$ and $\phi_{i}^{L}$ are realizations
of i.i.d uniformly distributed random variables $\Sigma$ and $\Phi^{L}$
over $(0,1)$ respectively, and letting the number of agents $n$
tend to infinity, we can express the consensus sequence $c_{k}=f(c_{k-1})\forall k\in\mathbb{N}$
as a quadratic and linear function of $c_{k-1}$ for $c_{k-1}\in[0,\frac{1}{2}]$
and $c_{k-1}\in(\frac{1}{2},1]$ respectively. 2) These expressions
will allow us to prove that for $X<\frac{1}{2}$, the consensus belief
sequence is monotonically increasing on $[0,1-X)$. 3) We will then
be able to conclude on the convergence of the sequence to $c_{\infty}=1-X$. 

We will first use the results from Proof of Proposition 4 to express
for all $k$ in $\mathbb{N}$ the sequence $c_{k}=f(c_{k-1})$ as
a quadratic and linear function of $c_{k-1}$ for $c_{k-1}\in[0,\frac{1}{2}]$
and $c_{k-1}\in(\frac{1}{2},1]$ respectively.

Recall that the consensus belief is defined as:

\[
c_{k}=\sum_{i=1}^{n}w_{i}\mu_{i,k},\forall k\in\mathbb{N},
\]

where $\mu_{i,k}$ represents the hatred belief of agent $i$ at timepoint
$k$ and $w_{i}$ is fixed $\forall i$ such that $\forall i\in n,w_{i}>0\mbox{ and }\sum_{i=1}^{n}w_{i}=1$. 

To explore the dynamics of the consensus belief sequence, we will
express $c_{k}$ as a function of the parameters $\sigma_{i},\phi_{i}^{L}$
for agents $i$, the constant $X$, and the previous consensus belief
$c_{k-1}$.

The Proof of Proposition 4 allows us to express $f(c_{k-1})$ over
the domain $c_{k-1}\in[0,\frac{1}{2}]$ as:
\begin{equation}
f(c_{k-1})=\sum_{i,\sigma_{i}\leq2Xc_{k-1}}\left[w_{i}\phi_{i}^{L}c_{k-1}\right]+\sum_{i,2Xc_{k-1}<\sigma_{i}\leq2X(1-c_{k-1})}\left[w_{i}c_{k-1}\right]+\sum_{i,\sigma_{i}>2X(1-c_{k-1})}\left[w_{i}\left(\phi_{i}^{L}+(1-\phi_{i}^{L})c_{k-1}\right)\right].\label{eq:ck_low}
\end{equation}

Similarly, over the domain $c_{k-1}\in(\frac{1}{2},1]$, we can write:
\begin{equation}
f(c_{k-1})=\sum_{i,\sigma_{i}\leq2X(\phi_{i}^{L}+c_{k-1}-2\phi_{i}^{L}c_{k-1})}\left[w_{i}\phi_{i}^{L}c_{k-1}\right]+\sum_{i,\sigma_{i}>2X(\phi_{i}^{L}+c_{k-1}-2\phi_{i}^{L}c_{k-1})}\left[w_{i}\left(\phi_{i}^{L}+(1-\phi_{i}^{L})c_{k-1}\right)\right].\label{eq:ck_high}
\end{equation}

We will now apply the assumption that $\sigma_{i}$ and $\phi_{i}^{L}$
are realizations of i.i.d uniformly distributed random variables $\Sigma$
and $\Phi^{L}$ respectively, to simplify Equations (\ref{eq:ck_low},\ref{eq:ck_high}). 

From this assumption, $c_{k}$ is the sum of i.i.d random variables
and is itself a random variable. We can thus apply the weak law of
large numbers as the number of agents $n\rightarrow\infty$. From
now on, we let $n\rightarrow\infty$. We can write: $c_{k}=\mathbb{E}(c_{k})$.
We will now show that this allows us to express the function $f$
over the domain $[0,\frac{1}{2}]$ and $(\frac{1}{2},1]$ as a quadratic
and linear function of $c_{k-1}$ respectively.

We will start by taking the expected values of $c_{k}$ as defined
in Equations (\ref{eq:ck_low},\ref{eq:ck_high}) respectively, and
simplify the expression for $f$ over $[0,1]$. In order to compute
these expected values, we will remove the dependencies on $\sigma_{i}$
of the sum terms in Equation (\ref{eq:ck_low},\ref{eq:ck_high})
and express $f$ as a single sum over all agents $i\in N$. To do
so, we let, $H$ be the Heaviside function defined over $\mathbb{R}$
such that:
\[
H(x):=\begin{cases}
0 & x\leq0\\
1 & x>0
\end{cases}.
\]

For clarity, we will compute the expected values of Equation (\ref{eq:ck_low})
and Equation (\ref{eq:ck_high}) separately.

We now consider $f$ in the region $c_{k-1}\in[0,\frac{1}{2}]$ defined
by Equation (\ref{eq:ck_low}) and make use of the Heaviside function
$H$ to compute the expected value for $c_{k}.$ Starting from Equation
(\ref{eq:ck_low}), we write:

\begin{align}
f(c_{k-1})= & \mathbb{E}\left[\sum_{i=1}^{n}w_{i}\left[\phi_{i}^{L}c_{k-1}+c_{k-1}(1-\phi_{i}^{L})H\left(\sigma_{i}-2Xc_{k-1}\right)+\phi_{i}^{L}(1-c_{k-1})H\left(\sigma_{i}-2X(1-c_{k-1})\right)\right]\right]\nonumber \\
= & \sum_{i=1}^{n}w_{i}\mathbb{E}\left[\Phi^{L}c_{k-1}+c_{k-1}(1-\Phi^{L})H\left(\Sigma-2Xc_{k-1}\right)+\Phi^{L}(1-c_{k-1})H\left(\Sigma-2X(1-c_{k-1})\right)\right]\label{eq:ck_expected_low}\\
= & \mathbb{E}\left[\Phi^{L}c_{k-1}+c_{k-1}(1-\Phi^{L})H\left(\Sigma-2Xc_{k-1}\right)+\Phi^{L}(1-c_{k-1})H\left(\Sigma-2X(1-c_{k-1})\right)\right].\nonumber 
\end{align}

To simplify Equation (\ref{eq:ck_expected_low}), we note that the
expected value of $H(\Sigma-a)$ for $a\in\mathbb{R}$ is equal to
the probability of the event $\Sigma>a$. When $a\in[0,1]$ and $\Sigma\sim\mathcal{U}(0,1)$
:
\[
\mathbb{E}\left[H(\Sigma-a)\right]=\mathbb{P}(\Sigma>a)=1-\mathbb{P}(\Sigma\leq a)=1-a
\]

We are now ready to express $f(c_{k-1})$ in the range $c_{k-1}\in[0,\frac{1}{2}]$
as a quadratic function of $c_{k-1}$ :
\begin{align}
f(c_{k-1})= & c_{k-1}\left(\frac{1}{2}+(1-\frac{1}{2})(1-2Xc_{k-1})-\frac{1}{2}\left(1-2X(1-c_{k-1})\right)\right)+\frac{1}{2}\left(1-2X(1-c_{k-1})\right)\label{eq:f_low}\\
= & -2Xc_{k-1}^{2}+(2X+\frac{1}{2})c_{k-1}+\frac{1}{2}-X.\nonumber 
\end{align}

Equation (\ref{eq:f_low}) will allow us to study the dynamics of
the consensus belief $c_{k}$ as $k$ increases in the range $c_{k-1}\in[0,\frac{1}{2}]$
. 

The next step of the analysis is to simplify the expression for $f$
as defined over the range $c_{k-1}\in(\frac{1}{2},1]$ by Equation
(\ref{eq:ck_high}). We will then be ready to explore the dynamics
of the consensus belief in the range $c_{k-1}\in[0,1]$. 

We will now express $f$ as defined in Equation (\ref{eq:ck_high})
as a linear function of $c_{k-1}$ by using the Heaviside function
and computing the expected value of $c_{k}$. 
\begin{align}
f(c_{k-1})= & \mathbb{E}\left[\sum_{i=1}^{n}w_{i}\left[\phi_{i}^{L}c_{k-1}+\left(\phi_{i}^{L}+(1-2\phi_{i}^{L})c_{k-1}\right)H\left(\sigma_{i}-2X(\phi_{i}^{L}+c_{k-1}-2\phi_{i}^{L}c_{k-1})\right)\right]\right]\nonumber \\
= & \sum_{i=1}^{n}w_{i}\mathbb{E}\left[\Phi^{L}c_{k-1}+\left(\Phi^{L}+(1-2\Phi^{L})c_{k-1}\right)H\left(\Sigma-2X(\Phi^{L}+c_{k-1}-2\Phi^{L}c_{k-1})\right)\right]\label{eq:ck_expected_high}\\
= & \mathbb{E}\left[\Phi^{L}c_{k-1}+\left(\Phi^{L}+(1-2\Phi^{L})c_{k-1}\right)H\left(\Sigma-2X(\Phi^{L}+c_{k-1}-2\Phi^{L}c_{k-1})\right)\right].\nonumber 
\end{align}

We must now compute the expected value expressed by Equation (\ref{eq:ck_expected_high}).
Specifically, we must calculate $\mathbb{E}\left[H\left(\Sigma-2X(\Phi^{L}+c_{k-1}-2\Phi^{L}c_{k-1})\right)\right]$.
To do so, we let $T$ be the random variable defined as:
\[
T=2X(c_{k-1}+\Phi^{L}(1-2c_{k-1})).
\]

We will now prove that $T$ is a uniformly distributed random variable
over $[2X(1-c_{k-1}),2Xc_{k-1}]$, which will allow us to prove that 

\[
\forall X\in[0,\frac{1}{2}),\forall c_{k-1}\in(\frac{1}{2},1],\mathbb{E}\left[H\left(\Sigma-2X(\Phi^{L}+c_{k-1}-2\Phi^{L}c_{k-1})\right)\right]=1-X.
\]

We first note that: 
\[
X=0,\Sigma\sim\mathcal{U}(0,1)\implies\mathbb{E}\left[H\left(\Sigma-2X(\Phi^{L}+c_{k-1}-2\Phi^{L}c_{k-1})\right)\right]=\mathbb{E}\left[H\left(\Sigma-0)\right)\right]=1=1-X.
\]

Then, we note that:
\[
X\in(0,\frac{1}{2}),c_{k-1}\in(\frac{1}{2},1]\implies\frac{dT}{d\Phi^{L}}=2X(1-2c_{k-1})<0.
\]

Thus $T$ is defined by a monotonically decreasing function of $\Phi^{L}$.
$\Phi^{L}\sim\mathcal{U}(0,1)$, therefore $T$ is uniformly distributed
over its range. The boundaries of $T$ are:
\begin{align*}
T(\Phi^{L}=0) & =2Xc_{k-1}.\\
T(\Phi^{L}=1) & =2X(1-c_{k-1}).
\end{align*}

We note that:
\[
X\in(0,\frac{1}{2}),c_{k-1}>\frac{1}{2}\implies2Xc_{k-1}>2X(1-c_{k-1}).
\]

Therefore, $T\sim\mathcal{U}(2X(1-c_{k-1}),2Xc_{k-1})$. By letting
$f_{T}(y)$ be the density function of the random variable $T$ we
write:
\[
f_{T}(y)=\begin{cases}
\frac{1}{2Xc_{k-1}-2X(1-c_{k-1})}=\frac{1}{2X(2c_{k-1}-1)}, & \mbox{if }y\in[2X(1-c_{k-1}),2Xc_{k-1}],\\
0, & \mbox{otherwise.}
\end{cases}
\]

We are now ready to compute $\mathbb{E}\left[H\left(\Sigma-2X(\Phi^{L}+c_{k-1}-2\Phi^{L}c_{k-1})\right)\right]$.
We let $F_{\Sigma}$ be the cumulative distribution function of the
random variable $\Sigma\sim\mathcal{U}(0,1)$ and write:
\begin{align}
\mathbb{E}\left[H(\Sigma-T)\right]= & 1-\mathbb{P}(\Sigma\leq T)\label{eq:exp_h_t}\\
= & 1-\int_{2X(1-c_{k-1})}^{2Xc_{k-1}}F_{\Sigma}(y)f_{T}(y)\mathrm{{d}y}\nonumber \\
= & 1-\frac{1}{2X(2c_{k-1}-1)}\int_{2X(1-c_{k-1})}^{2Xc_{k-1}}F_{\Sigma}(y)\mbox{dy}.\nonumber 
\end{align}

To calculate the integral in Equation (\ref{eq:exp_h_t}) we note
that 
\[
X\in[0,\frac{1}{2}),c_{k-1}\in[\frac{1}{2},1]\implies[2X(1-c_{k-1}),2Xc_{k-1}]\subseteq[0,1].
\]

Thus:
\begin{align*}
\mathbb{E}\left[H(\Sigma-T)\right]= & 1-\frac{1}{2X(2c_{k-1}-1)}\int_{2X(1-c_{k-1})}^{2Xc_{k-1}}y\mbox{dy}\\
= & 1-\frac{1}{2X(2c_{k-1}-1)}\cdot\frac{(2X)^{2}\cdot(c_{k-1}^{2}-(1-c_{k-1})^{2})}{2}\\
= & 1-\frac{2X(2c_{k-1}-1)}{2(2c_{k-1}-1)}\\
= & 1-X.
\end{align*}

Therefore we proved that:

\[
\forall X\in[0,\frac{1}{2}),\forall c_{k-1}\in(\frac{1}{2},1],\mathbb{E}\left[H\left(\Sigma-2X(\Phi^{L}+c_{k-1}-2\Phi^{L}c_{k-1})\right)\right]=1-X.
\]

We can now simplify the expression for $f$ defined in Equation (\ref{eq:ck_expected_high})
for $c_{k-1}\in(\frac{1}{2},1]$ as a linear function of $c_{k-1}$:
\begin{align}
f(c_{k-1})= & \mathbb{E}\left[\Phi^{L}c_{k-1}+\left(\Phi^{L}+(1-2\Phi^{L})c_{k-1}\right)H\left(\Sigma-2X(\Phi^{L}+c_{k-1}-2\Phi^{L}c_{k-1})\right)\right]\label{eq:f_high}\\
= & \frac{1}{2}c_{k-1}+(\frac{1}{2}+(1-2\frac{1}{2})c_{k-1})(1-X)\nonumber \\
= & \frac{1}{2}c_{k-1}+\frac{1-X}{2}.\nonumber 
\end{align}

Equation (\ref{eq:f_high}) will allow us to study the dynamics of
the consensus belief $c_{k}$ in the range $c_{k-1}\in(\frac{1}{2},1]$. 

The sequence of consensus belief $c_{k}$ is now defined by Equation
(\ref{eq:f_low}) and Equation (\ref{eq:f_high}) on $c_{k-1}\in[0,\frac{1}{2}]$
and $c_{k-1}\in(\frac{1}{2},1]$ respectively. This will allow us
to enter the second step of the proof and show that the sequence $c_{k}=f(c_{k-1})$
starting at $c_{0}=0$ is monotonically increasing on $c_{k-1}\in[0,1-X).$
We will then prove that the $\lim_{k\rightarrow\infty}c_{k}=1-X$. 

To prove that $c_{k}$ is monotonically increasing with $k$ on the
region $c_{k-1}\in[0,1-X)$, we will first prove that the consensus
belief increases montonically up to a value $c_{k^{*}}$ at a period
$k^{*}\in\mathbb{N}$ such that $c_{k^{*}}>\frac{1}{2}$. At that
timepoint, it will enter in the region where the sequence is defined
by Equation (\ref{eq:f_high}). We will then prove that it increases
monotonically and converges to $c_{\infty}=1-X$.

Let's prove that $\forall X\in[0,\frac{1}{2})$, the consensus belief
sequence is monotonically increasing in the range $c_{k-1}\in[0,\frac{1}{2}]$,
and reaches a value $c_{k^{*}}$ at a period $k^{*}\in\mathbb{N}$
such that $c_{k^{*}}>\frac{1}{2}$.

From Equation (\ref{eq:f_low}), we note that:
\begin{align}
c_{k}-c_{k-1} & =-2Xc_{k-1}^{2}+(2X+\frac{1}{2})c_{k-1}+\frac{1}{2}-X-c_{k-1}\label{eq:ck_diff}\\
 & =-2Xc_{k-1}^{2}+(2X-\frac{1}{2})c_{k-1}+\frac{1}{2}-X.\nonumber 
\end{align}

We will now prove that: 
\[
X\in[0,\frac{1}{2}),c_{k-1}\in[0,\frac{1}{2}]\implies c_{k}>c_{k-1}.
\]

First, we observe that for $X=0$:

\[
\forall c_{k-1}\in[0,\frac{1}{2}],X=0\implies c_{k}-c_{k-1}=\frac{1}{2}(1-c_{k-1})>0.
\]

Thus, for $X=0$, the sequence for $c_{k}$ is increasing for $c_{k-1}\in[0,\frac{1}{2}]$.
Let's now prove it is also the case for $X\in(0,\frac{1}{2})$.

We note that on $c_{k-1}\in[0,\frac{1}{2}]$:

\begin{equation}
c_{k}-c_{k-1}=0\iff-2Xc_{k-1}^{2}+(2X-\frac{1}{2})c_{k-1}+\frac{1}{2}-X=0.\label{eq:characteristic_small_c}
\end{equation}

We can now show that the quadratic function defined by Equation (\ref{eq:characteristic_small_c})
is striclty positive for $c_{k-1}\in[0,\frac{1}{2}]$, and thus that
the consensus sequence $c_{k}$ is monotonically increasing for $c_{k-1}\in[0,\frac{1}{2}]$.

When $X\in(0,\frac{1}{2})$, the discriminant of Equation (\ref{eq:characteristic_small_c})
is $\Delta=-4X^{2}+2X+\frac{1}{4}$, and $\Delta\in(\frac{1}{4},\frac{1}{2}]$.
Equation (\ref{eq:characteristic_small_c}) admits real solutions
$c_{1},c_{2}$ of the form

\[
c_{1}=\frac{(2X-\frac{1}{2})-\sqrt{-4X^{2}+2X+\frac{1}{4}}}{4X}\mbox{ and }c_{2}=\frac{(2X-\frac{1}{2})+\sqrt{-4X^{2}+2X+\frac{1}{4}}}{4X}.
\]

By substituting the extreme values for $\Delta,$ we observe that
$c_{1}<0$ and $c_{2}>\frac{1}{2}$. Therefore we observe that for
$c_{k-1}\in[0,\frac{1}{2}]$, $c_{k}-c_{k-1}$ is defined by a quadratic
function with a negative coefficient for the 2nd order polynomial
term, whose roots are $c_{1}<0$ and $c_{2}>\frac{1}{2}$. Therefore:
\[
X\in(0,\frac{1}{2}),c_{k-1}\in[0,\frac{1}{2}]\implies c_{k}-c_{k-1}>0.
\]

We proved that:
\[
X\in[0,\frac{1}{2}),c_{k-1}\in[0,\frac{1}{2}]\implies c_{k}>c_{k-1}.
\]

The sequence $c_{k}$ is thus monotonically incerasing for $c_{k-1}\in[0,\frac{1}{2}]$. 

Let's now show that the sequence of consensus belief $c_{k}$ will
reach a value $c_{k^{*}}$ at period $k^{*}$ such that $c_{k^{*}}>\frac{1}{2}$
with $c_{k^{*}-1}\leq\frac{1}{2}$. This will allow us to show that
the sequence enters the region $c_{k-1}>\frac{1}{2}$, in which it
will converge to $c_{\infty}=1-X$.

First, for $X=0$, we notice that 
\[
c_{k}=\frac{1}{2}-X+\frac{1}{2}c_{k-1}=\frac{1}{2}(1+c_{k-1})\implies c_{1}=\frac{1}{2}\implies c_{2}>\frac{1}{2}.
\]

Thus for $X=0$, the sequence of consensus belief $c_{k}$ reaches
a value $c_{k^{*}}$ at period $k^{*}\in\mathbb{N}$ such that $c_{k^{*}}>\frac{1}{2}$.

Let's prove that the statement holds $\forall X\in(0,\frac{1}{2})$.

For $c_{k-1}\in[0,\frac{1}{2}]$, $f$ is a polynomial function of
$c_{k-1}$ with a negative coefficient for the 2nd order polynomial
term. Therefore, $f$ is monotonically increasing on the region $(-\infty,\operatorname*{arg\,max}_{c_{k-1}}f(c_{k-1}))$
. We notice that:

\[
X\in(0,\frac{1}{2})\implies\operatorname*{arg\,max}_{c_{k-1}}f(c_{k-1})=\frac{1}{2}+\frac{1}{8X}>\frac{1}{2}.
\]

Thus $\forall X\in(0,\frac{1}{2})$, $f$ is monotonically increasing
on $c_{k-1}\in(-\infty,\frac{1}{2}+\frac{1}{8X})$, and specifically
in $[0,\frac{1}{2}]$. Thus:
\begin{equation}
\forall X\in(0,\frac{1}{2}),c_{k-1}\in[0,\frac{1}{2}]\implies\operatorname*{max}f(c_{k})=f(\frac{1}{2})=\frac{3}{4}-\frac{X}{2}>\frac{1}{2}.\label{eq:max_ck*}
\end{equation}

Therefore, for $c_{k}$ to remain smaller than $\frac{1}{2}$ for
all $k\in\mathbb{N}$, the sequence must converge to a point $c_{k-1}\in[0,\frac{1}{2}]$.
Recall that we proved Equation (\ref{eq:ck_diff}) admits no solution
in $c_{k-1}\in[0,\frac{1}{2}]$. Therefore, for $c_{k-1}\in[0,\frac{1}{2}]$
there exist no fixed point of $f$, and the consensus belief cannot
converge to a value $\in[0,\frac{1}{2}]$. Thus, because the sequence
of consensus beliefs $c_{k}$ is monotonically increasing for $c_{k-1}\in[0,\frac{1}{2}]$,
and that for $c_{k-1}\in[0,\frac{1}{2}]$, $\operatorname*{max}f(c_{k})>\frac{1}{2}$,
there exists a period $k^{*}\in\mathbb{N}$ such that $c_{k^{*}}>\frac{1}{2}$.

We will now prove that the consensus belief then increases monotonically
on $c_{k-1}\in(\frac{1}{2},1-X)$ and converges to $c_{\infty}=1-X$.

We will start by proving that the sequence of consensus belief is
monotonically increasing for $c_{k-1}\in(\frac{1}{2},1-X)$.

From Equation (\ref{eq:ck_high}), we note that:
\begin{equation}
c_{k}-c_{k-1}=\frac{1}{2}c_{k-1}+\frac{1-X}{2}-c_{k-1}=\frac{1-X}{2}-\frac{1}{2}c_{k-1}.\label{eq:ck_diff_high}
\end{equation}

Thus,
\begin{equation}
c_{k}-c_{k-1}>0\iff\frac{1-X}{2}-\frac{1}{2}c_{k-1}>0\iff c_{k-1}<1-X.\label{eq:cond_c_in_high}
\end{equation}

Therefore, we conclude by induction that $\forall k\in\mathbb{N},k\geq k^{*}\implies c_{k}<1-X$
and the sequence of consensus belief $c_{k}$ is monotonically increasing
on $c_{k-1}\in(\frac{1}{2},1-X]$.

We will now complete the proof by showing that $c_{k}$ is bounded
above by $1-X$, which allows us to prove the third step described,
namely that $\lim_{k\rightarrow\infty}c_{k}=1-X$. 

We use the result from Equation (\ref{eq:cond_c_in_high}) to prove
by induction that: 
\[
\forall k\in\mathbb{N},c_{k-1}\in(\frac{1}{2},1],c_{k-1}<1-X\implies c_{k}<1-X.
\]

Suppose that $c_{k-1}<1-X$, then:

\[
c_{k}<\frac{1}{2}(1-X)+\frac{1-X}{2}\implies c_{k}<1-X.
\]

We now verify that the relation holds for $c_{k^{*}}$, the first
term such that $c_{k}>\frac{1}{2}$. From Equation (\ref{eq:max_ck*}),
we note that $c_{k^{*}}$ is bounded such that:
\[
c_{k^{*}}\leq\frac{3}{4}-\frac{X}{2}.
\]

We observe that:
\[
\frac{3}{4}-\frac{X}{2}<1-X\iff X<\frac{1}{2}.
\]

Thus:
\[
X\in[0,\frac{1}{2})\implies c_{k^{*}}<1-X.
\]

The consensus sequence is thus bounded above by $1-X$. Therefore,
it is monotonically incereasing and bounded above, which implies its
limit $\lim_{k\rightarrow\infty}c_{k}$ exists. Let $c_{\infty}$
be this limit. We can express $c_{\infty}$ as:
\[
\lim_{k\rightarrow\infty}c_{k}=\lim_{k\rightarrow\infty}c_{k-1}\implies\frac{1}{2}c_{\infty}+\frac{1-X}{2}=c_{\infty}\implies c_{\infty}=1-X.
\]

We proved that, $\forall X\in(0,\frac{1}{2})$, by assuming $\sigma_{i}$
and $\phi_{i}^{L}$ are realizations of i.i.d uniformly distributed
random variables $\Sigma$ and $\Phi^{L}$ over $(0,1)$ respectively,
and letting the number of agents $n$ tend to infinity, the consensus
belief starting at $c_{0}=0$ is monotonically increasing with $k\in\mathbb{N}$
and converges to $c_{\infty}=1-X$.
\end{proof}
\pagebreak\textbf{Part 3}.\emph{ f is a piecewise linear function
of $M+1\in\mathbb{N}$ intervals $I_{1},...,I_{M+1}$, whose range
are defined by the $M$ discontinuities of $f$ over $[0,1]$. The
discontinuites repesent the unique threshold values of $c_{k-1}$
for which at least one agent $i$ changes the signal it samples. The
M+1 pairwise disjoint intervals are such that $\cup_{i=1}^{M+1}I_{i}=[0,1]$.
Some of these interval $I^{*}\subset I$ admit fixed points. If the
consensus belief converges to a fixed point, then it will converge
to the fixed point of the first interval it visits which belongs to
I{*}.}
\begin{proof}
We are interested in providing explicit equations for the fixed points
$c$ of each of the linear functions which define $f$ in its continuous
intervals. The discontinuities of $f$ repesent the unique threshold
values of $c_{k-1}$ for which at least one agent $i$ changes opinion.
To provide an explicit equation for each fixed point, we will first
proceed in three steps: 1) group the agents based on their change
in signal sampling as the consensus increases, and define $t_{i}$
as the threshold value of $c_{k-1}$ for agent $i$ for such a change
to occur, 2) order the agents in each of these respective groups such
that $i<j\implies t_{i}<t_{j}$, 3) express the equation for the fixed
point $c$ of each of these intervals. We will then prove that intervals
$I^{*}$ of $[0,1]$ which admit a fixed point are such that $f(I^{*})\subset I^{*}$.
This will allow us to prove that, if the consensus belief converges
to a fixed point $c$, then the fixed point is that of the linear
function defining $f$ in the first visited interval $I^{*}$.

We will start by providing explicit equations for the fixed points
of the linear functions which define each interval of $f$. 

From Proof of Proposition 5 part 2, we know that over $c_{k-1}\in[0,1]$,
$f$ is defined as a piecewise linear function expressed by Equations
(\ref{eq:ck_low},\ref{eq:ck_high}) for $c_{k-1}\in[0,\frac{1}{2}]$
and $c_{k-1}\in(\frac{1}{2},1]$ respectively. We will thus proceed
with our three step analysis for the range $c_{k-1}\in[0,\frac{1}{2}]$
and $c_{k-1}\in(\frac{1}{2},1]$ separately. 

Let's proceed with the three defined steps for the region $c_{k-1}\in[0,\frac{1}{2}]$.

1) We will show that we can group all agents in $i$ into three sub-groups
as $c_{k-1}$ increases in $[0,\frac{1}{2}]$: a) agents who always
sample the leader's signal, b) agents who stick to the consensus before
always sampling from the policy-maker when a threshold value for $c_{k-1}$
is reached, c) agents who stick to the consensus before always sampling
from the leader when a threshold value for $c_{k-1}$ is reached.

From Proof of Proposition 4, we recall that:

\[
\begin{cases}
0\leq\sigma_{i}\leq2Xc_{k-1}, & \mbox{the agent samples the policy-maker,}\\
2Xc_{k-1}<\sigma_{i}\leq2X(1-c_{k-1}), & \mbox{the agent sticks to the consensus,}\\
2X(1-c_{k-1})<\sigma_{i}\leq1, & \mbox{the agent samples the leader.}
\end{cases}
\]

To show that we can group the agents in a), b), and c), we first show
that at timepoint $k=1$, agents have either stuck to the consensus
or sampled from the leader:

\[
\forall i\in N,\sigma_{i}>0\mbox{ and }c_{0}=0\implies\sigma_{i}>2Xc_{0},
\]
and no agent sampled from the policy-maker. Hence at period $k=1$,
agents stuck to the consensus or sampled from the leader. This will
allow us to group the agents in the sub-groups as defined in a), b)
and c).

a) Some agents will always sample from the leader's signal as $c_{k-1}$
increases in $[0,\frac{1}{2}]$. Indeed, an agent $i$ will sample
the leader at period 1 if $\sigma_{i}>2X$. We notice that the threshold
$2X(1-c_{k-1})$ decreases as $c_{k-1}$ increases in $\mathbb{R}$
and escpecially on $[0,\frac{1}{2}]$. Thus, all agents $i$ such
that $\sigma_{i}>2X$ will sample from the leader for all $c_{k-1}\in[0,\frac{1}{2}]$
. Let $\mathcal{A}=\{i|\sigma_{i}>2X\}$. Without loss of generality,
we relabel the agents in $\mathcal{A}$ from $1$ to $|\mathcal{A}|=A$.

b) We will now show that some agents will change from sticking to
the consensus to sampling the policy-maker as $c_{k-1}$ increases
from $0$ to $\frac{1}{2}$. By observing that:

\[
\forall c_{k-1}\in[0,\frac{1}{2}],2X(1-c_{k-1})\in[X,2X],
\]
we state that:
\[
\forall c_{k-1}\in[0,\frac{1}{2}],\forall i,\sigma_{i}\leq X\implies\sigma_{i}\leq2X(1-c_{k-1}).
\]

Therefore, for $c_{k-1}\in[0,\frac{1}{2}]$, agents such that $\sigma_{i}\leq X$
will never sample the leader. In the limit $c_{k-1}\rightarrow\frac{1}{2},2Xc_{k-1}=2X(1-c_{k-1})$
and these agents will therefore change from sticking to the consenus
to sampling from the polic-maker as $c_{k-1}$ increases in $[0,\frac{1}{2}]$.
For each of these agents $i$, the change will occur at a threshold
value $t_{i}$ for $c_{k-1}$, such $t_{i}=\frac{\sigma_{i}}{2X}$.
Let $\mathcal{B}$ be the set of cardinality $B$ of agents $i$ such
that $\sigma_{i}\leq X$. Without loss of generality, we label the
agents in $\mathcal{B}$ from $A+1$ to $A+B$.

c) We will now show that some agents will change from sticking to
the consensus to sampling the leader as $c_{k-1}$ increases from
$0$ to $\frac{1}{2}$. By observing that:
\[
\forall c\in[0,\frac{1}{2}],2Xc\in[0,X],
\]
we state that:

\[
\forall c_{k-1}\in[0,\frac{1}{2}],\forall i,\sigma_{i}>X\implies\sigma_{i}>2Xc_{k-1}.
\]

Therefore, for $c_{k-1}\in[0,\frac{1}{2}]$, agents such that $\sigma_{i}>X$
will never sample from the policy-maker. Using the same argument as
in point b), as $c_{k-1}$ increases in $[0,\frac{1}{2}]$, these
agents will transfer from sticking to the consensus to sampling from
the leader. For each of these agents $i$, the change will occur at
a threshold $t_{i}=1-\frac{\sigma_{i}}{2X}$. Let $\mathcal{D}$ be
the set of cardinality $D$ of agents such that $\sigma_{i}>X$. Without
loss of generality, we label the agents in $\mathcal{D}$ from $A+B+1$
to $A+B+D$. 

Now that the agents who exhibit the same behaviour as $c_{k-1}$ increase
in $[0,\frac{1}{2}]$ are grouped, we are ready to reorder these agents
within each group. This will allow us to provide explicit expressions
for the fixed points $c$ of each of the linear functions which define
the intervals of $f$.

2) Without loss of generality, we re-order the agents in $\mathcal{B}\mbox{ and }\mathcal{D}$,
such that within each group $i<j\implies t_{i}\leq t_{j}$.

We will now express the explicit equations for the fixed points of
the linear functions defining $f$ for $c_{k-1}\in[0,\frac{1}{2}]$.

3) Let $\mathcal{E}$ be the set of agents who change opinion as $c_{k-1}$
increases in $[0,\frac{1}{2}]$ : $\mathcal{E}:=\mathcal{B}\cup\mathcal{D}$.
We order the set of threshold values $t_{i}$ for all $i$ in $\mathcal{E}$,
such that $i<j\implies t_{i}\leq t_{j}$. We have $E=|\mathcal{E}|=B+D$.
Let $I_{1},...,I_{B+D+1}$ be the $E$ pairwise disjoint intervals
such that $\cup_{i=1}^{B+D+1}I_{i}=[0,\frac{1}{2}]$, where $t_{1},...,t_{B+D}$
are the discontinuities of the function, $t_{0}=0$ and $t_{B+D+1}=\frac{1}{2}$.
We can now define the equation for the consensus update in the interval
$I_{m}=(t_{m},t_{m+1}]$, where $1\leq m\leq B+D$. We let $b$ be
the number of agents $i$ in $\mathcal{B}$ such that $t_{i}\leq t_{m}$,
and $d$ the number of agents $i$ in $\mathcal{D}$ such that $t_{i}\leq t_{m}$.
We define $f_{m}:=f(c_{k-1}),\forall c_{k-1}\in I_{m}$. According
to Equation (\ref{eq:ck_low}) we can express that:

\begin{align*}
f_{m}(c_{k-1}) & =\sum_{i=1}^{A}w_{i}(\phi_{i}^{L}+(1-\phi_{i}^{L})c_{k-1})+\sum_{i=A+1}^{A+b}w_{i}\phi_{i}c_{k-1}+\sum_{i=A+b+1}^{A+B}w_{i}c_{k-1}\\
 & +\sum_{i=A+B+1}^{A+B+d}w_{i}(\phi_{i}^{L}+(1-\phi_{i}^{L})c_{k-1})+\sum_{i=A+B+d+1}^{A+B+D}w_{i}c_{k-1}.
\end{align*}

We can now write an explicit equation for the fixed poins of the linear
intervals of $f$ for $c_{k-1}\in[0,\frac{1}{2}]$. A point $c$ must
be such that $c\in I_{m}$ and $f_{m}(c)=c$. We express the latter
condition as:

\begin{equation}
c=\frac{\sum_{i=1}^{A}w_{i}\phi_{i}^{L}+\sum_{i=A+B+1}^{A+B+d}w_{i}\phi_{i}^{L}}{1-\left[\sum_{i=1}^{A}w_{i}(1-\phi_{i}^{L})+\sum_{i=A+1}^{A+b}w_{i}\phi_{i}+\sum_{i=A+b+1}^{A+B}w_{i}+\sum_{i=A+B+1}^{A+B+d}w_{i}(1-\phi_{i}^{L})+\sum_{i=A+B+d+1}^{A+B+D}w_{i}\right]}.\label{eq:fp_small}
\end{equation}

We expressed the equation and conditions for a fixed point to exist
in the region $c_{k-1}\in[0,\frac{1}{2}]$. We will now repeat the
anlysis on the range $c_{k-1}\in(\frac{1}{2},1]$ before showing that
if these fixed points exist for $f$ over $c_{k-1}\in[0,1]$, then
they are stable and the sequence of consensus belief converges to
the fixed point $c$ of the first visited interval which admits a
fixed point.

Let's repeat the analysis for $f$ defined on $c_{k-1}\in(\frac{1}{2},1]$
by Equation (\ref{eq:f_high}).

1) We will start by showing that we can split the agents into four
distinct sub-groups as $c_{k-1}$ increases over $(\frac{1}{2},1]$:
a) agents who will always sample from the policy-maker, b) agents
who will always sample from the leader, c) agents who will transfer
from sampling the leader to sampling the policy-maker when a threshold
value for $c_{k-1}$ is reached, and d) agents who will transfer from
sampling the policy-maker to sampling the leader when a threshold
value for $c_{k-1}$ is reached. 

We will now understand the behaviour of agents within $c_{k-1}\in(\frac{1}{2},1]$,
to subsequently group them.

From Proof of Proposition 4, we can write:
\[
\begin{cases}
0\leq\sigma_{i}\leq2X(\phi_{i}^{L}+c_{k-1}-2\phi_{i}^{L}c_{k-1}), & \mbox{the agent samples the policy-maker},\\
2X(\phi_{i}^{L}+c_{k-1}-2\phi_{i}^{L}c)<\sigma_{i}\leq1, & \mbox{the agent samples the leader.}
\end{cases}
\]

Agents will change opinion if the order of $\sigma_{i}$ and $2X(\phi_{i}^{L}+c_{k-1}-2\phi_{i}^{L}c_{k-1})$
changes. To understand the behaviour of agents, we explore the variation
of $2X(\phi_{i}^{L}+c_{k-1}-2\phi_{i}^{L}c_{k-1})$ with $c_{k-1}$
in the range $(\frac{1}{2},1]$.
\begin{align}
\frac{\mathrm{d(2X(\phi_{i}^{L}+c_{k-1}-2\phi_{i}^{L}c_{k-1}))}}{\mathrm{dc_{k-1}}}= & 2X(1-2\phi_{i}^{L}).\label{eq:threshold_derivative}
\end{align}

The sign of $2X(1-2\phi_{i}^{L})$ varies on $\phi_{i}^{L}$, hence
we split the agents between $\phi_{i}^{L}\leq\frac{1}{2}$ and $\phi_{i}^{L}>\frac{1}{2}$
to analyse their respective decision making behaviour.

We will now explore the behaviour of each agent $i$as $c_{k-1}\mbox{ increases in }(\frac{1}{2},1]$
.

We first consider agents $i$ such that $\phi_{i}^{L}\leq\frac{1}{2}$
and observe that:
\[
\forall c_{k-1}\in(\frac{1}{2},1],\forall i,\phi_{i}^{L}\leq\frac{1}{2}\implies\frac{\mathrm{d(2X(\phi_{i}^{L}+c_{k-1}-2\phi_{i}^{L}c_{k-1}))}}{\mathrm{dc_{k-1}}}\geq0.
\]

We note that the thresholds at the boundaries of the interval $c_{k-1}\in(\frac{1}{2},1]$
for agents to change are respectively:
\begin{equation}
\begin{cases}
2X(\phi_{i}^{L}+\frac{1}{2}-2\frac{1}{2}\phi_{i}^{L})=X, & \mbox{if }c_{k-1}=\frac{1}{2},\\
2X(\phi_{i}^{L}+1-2\phi_{i}^{L})=2X(1-\phi_{i}^{L}), & \mbox{if }c_{k-1}=1.
\end{cases}\label{eq:boundaries_ck1}
\end{equation}

Thus, by using the bounds from Equation (\ref{eq:boundaries_ck1}),
we state that the threshold value of $\sigma$ for an agent to sample
the leader increase linearly from $X$ to $2X(1-\phi_{i}^{L})$ a
$c_{k-1}$ increases from $\frac{1}{2}$ to $1$. So we note that:
\[
\forall c_{k-}\in(\frac{1}{2},1],\forall i,\phi_{i}^{L}\leq\frac{1}{2},\begin{cases}
\sigma_{i}\leq X, & \implies\text{\mbox{agent \ensuremath{i} always samples the policy-maker, }}\\
\sigma_{i}>2X(1-\phi_{i}^{L}), & \implies\text{\mbox{agent \ensuremath{i} always samples the leader, }}\\
X<\sigma_{i}\leq2X(1-\phi_{i}^{L}), & \implies\mbox{agent \ensuremath{i} samples the leader, then the policy-maker.}
\end{cases}
\]

We now consider agents $i$ such that $\phi_{i}^{L}>\frac{1}{2}$.
We note that:
\[
\forall c_{k-1}\in(\frac{1}{2},1],\forall i,\phi_{i}^{L}>\frac{1}{2}\implies\frac{\mathrm{d(2X(\phi_{i}^{L}+c_{k-1}-2\phi_{i}^{L}c_{k-1}))}}{\mathrm{dc}_{k-1}}<0.
\]

Thus the threshold value of $\sigma_{i}$ for an agent to sample the
leader decreases linearly from $X$ to $2X(1-\phi_{i}^{L})$ as $c_{k-1}$
increases from $\frac{1}{2}$ to $1$. Therefore, we note that:
\[
\forall c_{k-1}\in(\frac{1}{2},1],\forall i,\phi_{i}^{L}>\frac{1}{2},\begin{cases}
\sigma_{i}\leq2X(1-\phi_{i}^{L}) & \implies\text{\mbox{agent \ensuremath{i} always samples the policy-maker. }}\\
\sigma_{i}>X & \implies\text{\mbox{agent \ensuremath{i} always samples the leader. }}\\
2X(1-\phi_{i}^{L})<\sigma_{i}\leq X & \implies\mbox{agent \ensuremath{i} samples the policy-maker, then the leader.}
\end{cases}
\]

We are now ready to split the population into the four distinct sets
defined in a), b), c) and d). From these groups, we will be able to
write down the expressions for fixed points of the affine functions
which define $f$ in the region $c_{k-1}\in(\frac{1}{2},1]$.

Let's group the agents.

a) Agents who will always sample from the policy-maker. We showed
that agents $i$ such that 
\[
\begin{cases}
\phi_{i}^{L}\leq\frac{1}{2}, & \sigma_{i}\leq X,\\
\phi_{i}^{L}>\frac{1}{2}, & \sigma_{i}\leq2X(1-\phi_{i}^{L}).
\end{cases}
\]

always sample from the policy-maker. We denote $\mathcal{R}$ the
set of such agents, with $R=|\mathcal{R}|$. Without loss of generality,
we label the agents in $\mathcal{R}$ from $1$ to $R$.

b) Agents who will always sample from the leader. We showed that agents
$i$ such that 
\[
\begin{cases}
\phi_{i}^{L}\leq\frac{1}{2}, & \sigma_{i}>2X(1-\phi_{i}^{L}),\\
\phi_{i}^{L}>\frac{1}{2}, & \sigma_{i}>X.
\end{cases}
\]

always sample from the leader. We denote $\mathcal{U}$ the set of
such agents, with $U=|\mathcal{U}|$. Without loss of generality,
we label the agents in $\mathcal{U}$ from $R+1$ to $R+U$.

c) Agents who will transfer from sampling the leader to sampling the
policy-maker when a threshold value for $c_{k-1}$ is reached. We
showed these agents $i$ are such that 

\[
\phi_{i}^{L}\leq\frac{1}{2},X<\sigma_{i}\leq2X(1-\phi_{i}^{L}).
\]

Let $\mathcal{W}$ be the set of such agents $i$, with $W=|\mathcal{W}|$.
These agents will change opinon when threshold values $t_{i}$ of
$c_{k-1}$ is reached such that the order between $\sigma_{i}$ and
$2X(\phi_{i}^{L}+c_{k-1}-2\phi_{i}^{L}c_{k-1})$ changes. Without
loss of generality, we label the agents in $\mathcal{W}$ from $R+U+1$
to $R+U+W$. We define $\forall i\in\mathcal{W},t_{i}=\frac{\frac{\sigma_{i}}{2X}-\phi_{i}^{L}}{1-2\phi_{i}^{L}}$.

d) Agents who will transfer from sampling the policy-maker to sampling
the leader when a threshold value for $c_{k-1}$ is reached. We showed
these agents $i$ are such that 

\[
\phi_{i}^{L}>\frac{1}{2},2X(1-\phi_{i}^{L})<\sigma_{i}\leq X.
\]

Let $\mathcal{Y}$ be the set of such agents $i$, with $Y=|\mathcal{Y}|$.
These agents will change opinon when threshold values $t_{i}$ of
$c_{k-1}$ is reached such that the order between $\sigma_{i}$ and
$2X(\phi_{i}^{L}+c_{k-1}-2\phi_{i}^{L}c_{k-1})$ changes. Without
loss of generality, we label the agents in $\mathcal{Y}$ from $R+U+W+1$
to $R+U+W+Y$. We define $\forall i\in\mathcal{Y},t_{i}=\frac{\frac{\sigma_{i}}{2X}-\phi_{i}^{L}}{1-2\phi_{i}^{L}}$.

We grouped agents into four distinct sub-groups. We are now ready
to reorder the agents within the groups and express the fixed points
of each of the linear functions defining $f$ for $c_{k-1}\in(\frac{1}{2},1]$.

2) Without loss of generality, we re-order the agents in $\mathcal{W}\mbox{ and }\mathcal{Y}$,
such that within each group $i<j\implies t_{i}\leq t_{j}$.

We will now express the explicit equations for the fixed points of
the linear functions defining $f$ for $c_{k-1}\in(\frac{1}{2},1]$.

3) Let $\mathcal{Z}$ be the set of agents who change opinion as $c_{k-1}$
increases in $(\frac{1}{2},1]$: $\mathcal{Z}:=\mathcal{W}\cup\mathcal{Y}$.
We order the set of threshold values $t_{i}$ for all $i$ in $\mathcal{Z}$,
such that $i<j\implies t_{i}\leq t_{j}$. We have $Z=|\mathcal{Z}|=W+Y$.
Let $I_{B+D+2},...,I_{B+D+W+Y+2}$ be the $Z$ pairwise disjoint intervals
such that $\cup_{i=B+D+2}^{B+D+W+Y+2}I_{i}=(\frac{1}{2},1]$, where
$t_{B+D+2},...,t_{B+D+W+Y+1}$ are the discontinuities of $f$ on
$c_{k-1}\in(\frac{1}{2},1]$, and $t_{B+D+W+Y+1}=1$. We can now define
the equation for the consensus update in the interval $I_{m}=(t_{m},t_{m+1}]$,
where $B+D+2\leq m<B+D+W+Y+2$. We let $w$ be the number of agents
$i$ in $\mathcal{W}$ such that $t_{i}\leq t_{m}$, and $y$ the
number of agents $i$ in $\mathcal{Y}$ such that $t_{i}\leq t_{m}$.
We define $f_{m}:=f(c_{k-1}),\forall c_{k-1}\in I_{m}$. According
to Equation (\ref{eq:ck_high}) we can express that:
\begin{align*}
f_{m}(c_{k-1})= & \sum_{i=1}^{R}w_{i}\phi_{i}^{L}c_{k-1}+\sum_{i=R+1}^{R+U}w_{i}(\phi_{i}^{L}+(1-\phi_{i}^{L})c_{k-1})+\sum_{i=R+U+1}^{R+U+w}w_{i}\phi_{i}^{L}c_{k-1}+\sum_{i=R+U+w+1}^{R+U+W}w_{i}(\phi_{i}^{L}+(1-\phi_{i}^{L})c_{k-1})\\
 & +\sum_{i=R+U+W+1}^{R+U+W+y}w_{i}(\phi_{i}^{L}+(1-\phi_{i}^{L})c_{k-1})+\sum_{i=R+U+W+y+1}^{R+U+W+Y}w_{i}\phi_{i}^{L}c_{k-1}
\end{align*}

To be a fixed point in the interval $m$, a point $c$ must be such
that $c\in I_{m}$ and $f_{m}(c)=c$.

\begin{equation}\label{eq:fp_high}
c=\dfrac{ \sum\limits_{R+1}^{R+U}w_i\phi_i^L + \sum\limits_{P+U+w+1}^{P+U+W+y}w_i\phi_i^L}{ 1 - \left[ \sum\limits_{i=1}^{R}w_i\phi_i^L + \sum\limits_{i=R+1}^{R+U}w_i(1-\phi_i^L) + \sum\limits_{R+U+1}^{P+U+w}w_i\phi_i^L+ \sum\limits_{R+U+w+1}^{R+U+W+y}w_i(1-\phi_i^L) + \sum\limits_{R+U+W+y+1}^{R+U+W+Y}w_i\phi_i^L  \right]}
\end{equation}

We expressed the equation and conditions for a fixed point to exist
in the region $c_{k-1}\in[0,1]$. We are now ready to study the dynamics
of these fixed points, and conclude on the convergence of the sequence
of consensus belief.

We have shown how to compute the fixed ponits for the affine function
defined within each interval separating discontinuities. We will now
prove that such fixed points which are in $f$ are stable.

With $M+1$ as the number of distinct linear intervals of $f$ for
$c_{k-1}\in[0,1]$, $\forall m\in[1,...,M+1]$, we define the slope
of the affine function in $I_{m}$ as $\beta_{m}$. If we define $n_{1m},n_{2m},n_{3m}$
as the sets of indices of agents who choose to sample the policy-maker,
stick to the consensus, and sample the leader respectively in each
interval $I_{m}$, we can write:

\[
\forall m,\beta_{m}=\sum_{i\in n_{1m}}w_{i}\phi_{i}^{L}+\sum_{i\in n_{2m}}w_{i}+\sum_{i\in n_{3m}}w_{i}(1-\phi_{i}^{L}).
\]

We note that:
\[
\forall i\in n,\phi_{i}^{L}\in(0,1),w_{i}\in(0,1),\sum_{i=1}^{n}w_{i}=1.
\]

Thus with $n_{1m}\cup n_{2m}\neq\emptyset$,
\[
0<\sum_{i\in n_{1m}}w_{i}\phi_{i}^{L}+\sum_{i\in n_{2m}}w_{i}+\sum_{i\in n_{3m}}w_{i}(1-\phi_{i}^{L})<\sum_{i\in n_{1m}}w_{i}+\sum_{i\in n_{2m}}w_{i}+\sum_{i\in n_{3m}}w_{i}=1
\]

Thus, for any continuous interval $I_{m}$ of the function, if a fixed
point $c\in I_{m}$, $|\beta_{m}|<1$ and $F(I_{m})\subseteq I_{m}$.
Therefore, all fixed point defined over $[0,1]$ are stable.

Thus, if a fixed point $c$ exists in $c_{k-1}\in[0,1]$, the fixed
point is a solution of Equation (\ref{eq:fp_small}) or Equation (\ref{eq:fp_high})
if $c$ is such that $c\leq\frac{1}{2}$ or $c>\frac{1}{2}$ respectively.
We have proved that if the consensus belief visits an interval $I_{m}$
such that $f$ admits a fixed point on $I_{m}$ , then the consensus
belief $c_{k}$ will converge to that fixed point as $k\rightarrow\infty$.
The consensus belief can therefore admit a fixed point and converge
to a single value.
\end{proof}
\pagebreak\textbf{Part 4. }\emph{Let $I_{v}$ be the set of intervals
visited by the consensus belief sequence $c_{k}=f(c_{k-1})$ as $k\rightarrow\infty$.
Let $x\in\mathbb{N^{*}}$. Let $D_{x}$ be the set of discontinuities
of the $x$th iterate of $f$ over $c_{k-1}\in[0,1]$, defined as
$f^{x}$. If $\nexists I^{*}\in I_{v},f(I^{*})\subset I^{*},$ and
$\exists x^{*}\in\text{\ensuremath{\mathbb{N}}},\forall x\geq x^{*},\forall c_{k-1}\in[0,1],f^{x}(c_{k-1})\in D_{x-1}\implies c_{k-1}\in D_{x-1}$,
then the consensus belief sequence is eventually periodic. Once the
consensus belief is periodic, the agents are split into three distinct
sub-groups: a) agents who will always sample the policy-maker, b)
agents who will always sample the leader, c) agents who will periodically
oscillate between a combination of sampling the leader, the policy-maker,
or sticking to the consensus. }
\begin{proof}
We will prove Proposition 4 by showing that if the discontinuities
in the iterates $f^{x}$ of function $f$ are finite as $x\rightarrow\infty$,
then $f$ is a piecewise linear function whose iterates converge to
a piecewise step function. The discontinuities of the piecewise step
function are a subset of the discontinuities of all the iterates of
$f$. If the consensus belief visits no interval of $f$ admitting
a fixed point, then the sequence of consensus belief is eventually
periodic. 

We sart by expressing a condition for the disconinuities in the iterates
$f^{x}$ of $f$ to be finite as $x\rightarrow\infty$. This will
allow us to prove that $\lim_{x\rightarrow\infty}f^{x}$ is a stepwise
function with a fintite number of discontinuities, and thus show that
the sequence of consensus belief is eventually periodic. 

Let $x\in\mathbb{N}$ and $f^{x}$ be the $x$th iterate of $f$.
We define $D_{x}$ as the set of discontinuities of $f^{x}$ over
$c_{k-1}\in[0,1]$. Assume that:
\begin{equation}
\exists x^{*}\in\text{\ensuremath{\mathbb{N}}},\forall x\geq x^{*},\forall c_{k-1}\in[0,1],f^{x}(c_{k-1})\in D_{x-1}\implies c_{k-1}\in D_{x-1}.\label{eq:disc_cond}
\end{equation}

Condition (\ref{eq:disc_cond}) ensures that there exists $x^{*}\in\mathbb{N}$,
such that from the iterate $x^{*}$ of $f$ onwards, no new pre-image
$c_{k-1}\in[0,1]$ yield a discontinuity. Therefore, there is a maximum
and finite number of disconinuities for all iterates $f^{x},x>x^{*}$. 

From this statement, we will prove that the consensus sequence $c_{k}$
is eventually periodic by showing that $f$ tends to a piecewise step
function where each continuous interval converges to a value in $[0,1]$.
As the number of values visited by the consensus $c_{k}$ would then
become finite, the sequence of consensus belief will become periodic.

To do so, we let $g$ be the first iterate of $f$ for which Condition
(\ref{eq:disc_cond}) holds. Suppose there are $M-1$ discontinuities
in the function $g$, $(M-1)\in\mathbb{N^{*}}$. We have shown in
Proof of Statement 3 of Proposition 5 that $f$ is a piecewise linear
function, where each affine interval $m$ is defined with a slope
$\beta_{m}$, with $\beta_{m}<1$. We also let $\gamma_{m}$ be the
Y-intercept of each affine function $m$ defining $f$. We can write:
\[
\forall m\in[1,...,M],g_{m}(c_{k-1})=\beta_{m}c_{k-1}+\gamma_{m}.
\]

Writing $g_{m}^{2}(c_{k-1})$ as the second iterate of the function
$g$ over each interval $m$, we have:

\begin{align*}
g_{m}^{2}(c_{k-1})= & \beta_{m}(\beta_{m}c_{k-1}+\gamma_{m})+\gamma_{m}\\
= & \beta_{m}^{2}c_{k-1}+\beta_{m}\gamma_{m}+\gamma_{m}
\end{align*}

Generalising for all $p\in\mathbb{N}$:

\begin{equation}
g_{m}^{p}(c_{k-1})=\beta_{m}^{p}c_{k-1}+\sum_{i=0}^{p-1}\beta_{m}^{i}\gamma_{m}\label{eq:gm_p}
\end{equation}

We must now prove that $\lim_{n\rightarrow\infty}g_{m}^{n}(c_{k-1})$
converges for any $m$ in $[1,...,M]$. Equation (\ref{eq:gm_p})
implies that:

\[
\beta_{m}g_{m}^{p}=\beta_{m}^{p+1}c_{k-1}+\sum_{i=1}^{p}\beta_{m}^{i}\gamma_{m}
\]

Therefore,
\begin{align*}
\beta_{m}g_{m}^{p}-g_{m}^{p}= & \beta_{m}^{p}c_{k-1}(\beta_{m}-1)+\beta_{m}^{p}-\gamma_{m}\\
\implies g_{m}^{p}= & \frac{\beta_{m}^{p}\left[c_{k-1}(\beta_{m}-1)+1\right]-\gamma_{m}}{\beta_{m}-1}
\end{align*}

Following the same reasoning as in in Proof of Statement 3 of Proposition
5, as the iterates of $f$ represent a new decision making step of
the agents in the population, we must have that the slope $\beta_{m}<1$
for all interval $m$ in $g(c_{k-1})$. Thus, $0<\beta_{m}<1$, and
at infinite horizon we have:

\[
\lim_{p\rightarrow\infty}g_{m}^{p}(c_{k-1})=\frac{\gamma_{m}}{1-\beta_{m}}
\]

To show that $\lim_{p\rightarrow\infty}g_{m}^{p}(c_{k-1})$ converges
for all $\forall m\in[1,...,M]$, we define $n_{1m},n_{2m},n_{3m}$
as the sets of agents who choose the policy-maker, consensus and leader
respectively, in interval $m$, and write:
\[
\forall m\in[1,...,M],\beta_{m}=\sum_{i\in n_{1m}}w_{i}\phi_{i}^{L}+\sum_{i\in n_{2m}}w_{i}+\sum_{i\in n_{3m}}w_{i}(1-\phi_{i}^{L}).
\]

Correspondingly we have:
\[
\forall m,\gamma_{m}=\sum_{i\in n_{3m}}w_{i}\phi_{i}^{L}.
\]

Therefore:
\begin{align*}
\forall m\in[1,...,M],\beta_{m}+\gamma_{m}= & \sum_{i\in n_{1m}}w_{i}\phi_{i}^{L}+\sum_{i\in n_{2m}}w_{i}+\sum_{i\in n_{3m}}w_{i}(1-\phi_{i}^{L})+\sum_{i\in n_{3m}}w_{i}\phi_{i}^{L}\\
= & \sum_{i\in n_{1m}}w_{i}\phi_{i}^{L}+\sum_{i\in n_{2m}}w_{i}+\sum_{i\in n_{3m}}w_{i}\\
< & \sum_{i\in n_{1m}}w_{i}+\sum_{i\in n_{2m}}w_{i}+\sum_{i\in n_{3m}}w_{i}\\
= & 1.
\end{align*}

Thus,
\[
\forall m\in[1,...,M],\beta_{m}+\gamma_{m}<1\text{\ensuremath{\implies}\ensuremath{\gamma_{m}<1-\beta_{m},}}
\]
and with $\gamma_{m}\in(0,1),\beta_{m}\in(0,1)\forall m\in[1,...,M]$,

\[
\forall c_{k-1}\in[0,1],\lim_{p\rightarrow\infty}g_{m}^{p}(c_{k-1})\in(0,1).
\]

Therefore, each affine interval $m$ converges when $p$ tends to
$\infty$, and the function $g^{p}$ tends to a piecewise step function
of $M$ constant values $\in(0,1)$.

We will now show that a sequence defined by a function $g^{p}$ which
is a piecewise step function of $M$ constant values is eventually
periodic.

Let:
\[
\forall m\in[1,...,M],l_{m}=\lim_{p\rightarrow\infty}g_{m}^{p}(c_{k-1}).
\]

In the limit where $p\rightarrow\infty$:
\[
g^{p}([0,1])=\{l_{1},...,l_{M}\}.
\]

The sequence becomes the mapping of a finite set within itself, which
is eventually periodic. This completes the proof of statement 3.
\end{proof}
We will now analyse the behaviour of agents when the dynamical system
is periodic.

Assume that the dynamical system of the consensus belief is periodic,
with $p_{0}$ and $p_{1}$ the extremum values of the period, such
that $0<p_{0}<p_{1}\leq1$. the behaviour of the agents can be separated
into: 1) agents who do not change opinion as the system evolves over
the period, 2) agents who oscillate between different opinions. We
consider the three cases for the values of $p_{0}$ and $p_{1}$,
namely a) $0<p_{0}<p_{1}\leq\frac{1}{2}$, b) $0<p_{0}\leq\frac{1}{2}\mbox{ and }\frac{1}{2}<p_{1}\leq1$,
c)$\frac{1}{2}<p_{0}<p_{1}\leq1$.

\textbf{Case a. $0<p_{0}<p_{1}\leq\frac{1}{2}$}

In this range of periodic orbit, the agents' behaviours are split
as follows:
\[
\begin{cases}
\sigma_{i}\leq2Xp_{0}, & \mbox{the agent will always sample from the policy-maker,}\\
\sigma_{i}>2X(1-p_{0}), & \mbox{the agent will always sample from the leader},\\
2Xp_{0}<\sigma_{i}\leq2Xp_{1}, & \mbox{the agent will oscillate between the consensus and the policy-maker,}\\
2X(1-p_{1})<\sigma_{i}\leq2X(1-p_{0}), & \mbox{the agent will oscillate between the consensus and the leader}.
\end{cases}
\]

\textbf{Case b. $0<p_{0}\leq\frac{1}{2}\mbox{ and }\frac{1}{2}<p_{1}\leq1$}

In this range of extremum values, the agents' behaviours are split
as follows:

For agents $i$ such that $\phi_{i}^{L}<\frac{1}{2}$:
\[
\sigma_{i}\leq2Xp_{0}\mbox{ and }\sigma_{i}\leq2X(\phi_{i}^{L}+p_{1}-2\phi_{i}^{L}p_{1})\implies\mbox{the agent will always sample from the policy-maker}
\]

For agents $i$ such that $\phi_{i}^{L}>\frac{1}{2}$:
\[
\sigma_{i}>2X(1-p_{0})\mbox{ and }\sigma_{i}>2X(\phi_{i}^{L}+p_{1}-2\phi_{i}^{L}p_{1})\implies\mbox{the agent will always sample from the leader}
\]

For agents $i$ such that $\phi_{i}^{L}=\frac{1}{2}$:
\[
\begin{cases}
\sigma_{i}\leq2Xp_{0}, & \mbox{the agent will always sample from the policy-maker,}\\
\sigma_{i}>2X(1-p_{0}), & \mbox{the agent will always sample from the leader}.
\end{cases}
\]

Remaining agents will oscillate either from consensus to policy-maker
or leader, or between leader and policy-maker.

\textbf{Case c. $\frac{1}{2}<p_{0}<p_{1}\leq1$}

In this range of extremum values, the agents' behaviours are split
as follows:

For agents $i$ such that $\phi_{i}^{L}\leq\frac{1}{2}$:
\[
\begin{cases}
\sigma_{i}\leq2X(\phi_{i}^{L}+p_{0}-2\phi_{i}^{L}p_{0}), & \mbox{the agent will always sample from the policy-maker,}\\
\sigma_{i}>2X(\phi_{i}^{L}+p_{1}-2\phi_{i}^{L}p_{1}), & \mbox{the agent will always sample from the leader}.
\end{cases}
\]

For agents $i$ such that $\phi_{i}^{L}>\frac{1}{2}$:
\[
\begin{cases}
\sigma_{i}\leq2X(\phi_{i}^{L}+p_{1}-2\phi_{i}^{L}p_{1}), & \mbox{the agent will always sample from the policy-maker,}\\
\sigma_{i}>2X(\phi_{i}^{L}+p_{0}-2\phi_{i}^{L}p_{0}), & \mbox{the agent will always sample from the leader}.
\end{cases}
\]

Remaining agents will oscillate between sampling from the policy-maker
and the leader.
\end{document}